\documentstyle[11pt,epsfig]{article}
\newcommand \be{\begin{equation}}
\newcommand \ba{\begin{eqnarray}}
\newcommand \ee{\end{equation}}
\newcommand \ea{\end{eqnarray}}

[1999/03/29 PSNFSS v.7.2
Times font as default roman
: S Rahtz]

\begin{document}

\begin{center}
{\LARGE Statistical Significance of Periodicity and
Log-Periodicity \\with Heavy-Tailed Correlated Noise}
\end{center}
\bigskip
\begin{center}
{\large Wei-Xing Zhou {\small$^{\mbox{\ref{igpp}}}$} and Didier
Sornette {\small$^{\mbox{\ref{igpp},\ref{ess},\ref{lpec}}}$}}
\end{center}
\bigskip
\begin{enumerate}

\item Institute of Geophysics and Planetary Physics, University of
California, Los Angeles, CA 90095\label{igpp}

\item Department of Earth and Space Sciences, University of
California, Los Angeles, CA 90095\label{ess}

\item Laboratoire de Physique de la Mati\`ere Condens\'ee, CNRS UMR
6622 and Universit\'e de Nice-Sophia Antipolis, 06108 Nice Cedex
2, France\label{lpec}

\end{enumerate}

%%\date{today}

\begin{abstract}
We estimate the probability that random noise, of several
plausible standard distributions, creates a false alarm that a periodicity
(or log-periodicity) is found in a time series. The solution of this
problem is already known for independent Gaussian distributed noise.
We investigate more general situations with non-Gaussian correlated noises
and present synthetic tests on the detectability and statistical
significance of periodic components. A periodic component of a time series
is usually detected by some sort of Fourier analysis. Here, we use
the Lomb periodogram analysis which is suitable and outperforms 
Fourier transforms
for unevenly sampled time series. We examine the
false-alarm probability of the largest spectral peak of the Lomb
periodogram in the presence of
power-law distributed noises, of short-range and of long-range
fractional-Gaussian noises. Increasing heavy-tailness (respectively
correlations describing persistence) tends to decrease
(respectively increase) the false-alarm probability of finding
a large spurious Lomb peak. Increasing
anti-persistence tends to decrease the false-alarm probability. We
also study the interplay between heavy-tailness and long-range
correlations.  In order to fully determine if a Lomb peak
signals a genuine rather than a spurious periodicity, one should in principle
characterize the Lomb peak height, its width and its
relations to other peaks in the complete spectrum.
As a step towards this full characterization, we construct the 
joint-distribution
of the frequency position (relative to other peaks)
and of the height of the highest peak of the power spectrum. We also
provide the distributions of the ratio of the second
highest Lomb peak to the maximum peak. Using the insight obtained
by the present statistical study, we
re-examine previously reported claims of ``log-periodicity'' and find
that the credibility for log-periodicity in 2D-freely decaying turbulence is
weakened while it is strengthened
for fracture, for the ion-signature prior to the Kobe earthquake and for
financial markets.

\end{abstract}

\section{Introduction}
\label{sec:intro}

The problem of detecting oscillatory components in noisy data is
one of the most general problems found in all scientific
disciplines. Without being exhaustive, the search of periodic
signals in complex time series goes from astrophysics (for
instance in the detection of planets outside our solar system),
geosciences (for instance in the detection of cycles in
meteorology and climate dynamics), medicine (for instance in the
effect of circadian cycles), to economics (for instance in the
business cycles or in the calendar effects occurring at intra-day,
week, month and year time scales in financial times series). It is
not stretching reality to state that almost all fields have at
least one outstanding question related to the detection of tiny
periodic oscillations in an otherwise noisy and complex
background. Many techniques have been invented, adapted or
improved to address the specificities of each situation.

The most standard tool for detecting a periodic component in a
noisy time series is the Fourier transform (FT) and, when
possible, the fast Fourier transform (FFT).

The FT and FFT methods require evenly sampled signals. However,
due to measurement constraints, it can happen that the time series
is not evenly spaced in time. A straightforward approach is to
reconstruct evenly spaced data using interpolation or rebinning
and then apply the FFT method. However, interpolation and
rebinning both introduce some distortion and may lose some
information in situations where the periodic signal is weak and
difficult to distinguish from noise.

In addition, the FT and FFT methods perform well only when the
the data size is relatively large so as to minimize aliasing (window border)
distortions at the extremities of the time series. Wavelet
techniques and singular spectral analysis have been developed to
address in part this problem.

A completely different method for detecting periodic components in
an unevenly sampled and relatively short signal is the Lomb
periodogram \cite{Press1996}. The Lomb periodogram analysis
performs local least-squares fit of the data by sinusoids centered
on each data point of the time series \cite{Scarg1982}. Given a
unevenly spaced signal $y(t_i)$, where $i=1, 2, \cdots, N$, the
normalized Lomb periodogram is calculated according to the
following formula \cite{Scarg1982,Horne1986,Press1996}:
\begin{equation}
   P_N(\omega) = \frac{1}{2\sigma^2}
\left[\frac{[\sum_{i=1}^N
y(t_i)\cos\omega(t_i-\tau)]^2}{\sum_{i=1}^N
\cos^2\omega(t_i-\tau)}
        +\frac{[\sum_{i=1}^N y(t_i)\sin\omega(t_i-\tau)]^2}
        {\sum_{i=1}^N \sin^2\omega(t_i-\tau)} \right]~,
\label{lombdefspec}
\end{equation}
where $\tau$ is defined by the equation
\begin{equation}
   \tan(2\omega\tau) = \left(\sum_{i=1}^N \sin 2
    \omega t_i \right) / \left(\sum_{i=1}^N \cos 2 \omega t_i \right),
\end{equation}
and $\sigma^2$ is the total variance of the data (including noise
and signal)\footnote{The formula (\ref{lombdefspec}) uses the
total variance of the data which is generally available rather
than the variance of the noise derived either from the residuals
after subtracting a signal or from the uncertainty of the
measurements \cite{Horne1986}.}. The normalized Lomb periodogram
$P_N(\omega)$ is similar to a FT power spectrum in which the
presence of peaks at certain angular frequencies $w_n$ signals the
possible existence of periodic components. The largest peaks point
to the most important frequencies in the time series, the higher
is a peak, the larger is the statistical significance of the
corresponding periodic component. Due to its local least-square
fit nature, the Lomb periodogram analysis works equally well for
unevenly spaced time series. It is also much less prone to
aliasing (window border) distortions in short time series.

To understand intuitively how the expression (\ref{lombdefspec})
allows one to detect periodic components, let us consider a noisy
signal $y(t)$ made of a finite number of independent periodic
functions plus a noise $x(t)$ of zero mean which is independent of
the periodic components: \be y(t) = \sum_{m=1}^M y_m \cos(\omega_m
t) + x(t)~. \label{Eq:yt} \ee Using the independence between the
noise and the periodic signals, we obtain $\langle y \rangle = 0$
and $\sigma^2 = \langle y^2 \rangle = \sum_{1}^M y_m^2 / 2 +
\sigma_x^2$. Since the translation $\tau$ in
Eq.~(\ref{lombdefspec}) provides usually a very minor correction,
we can ignore it here. In addition, we assume that the $t_i$'s are
not badly bunched \cite{Horne1986}. Given these two conditions,
the Lomb periodogram at angular frequency $\omega_m$ can be
written as
\begin{equation}
P_N(\omega_m) = \frac{N}{2} \left( \frac{\sum_{1}^M y_m^2}{y_m^2}
+ \frac{2\sigma_x^2}{y_m^2} \right)^{-1}~.
\label{Eq:PNapprox}
\end{equation}
This result (\ref{Eq:PNapprox}) emphasizes the importance of
normalizing the periodogram by the total variance of the data.
It also shows that the Lomb peak height is approximately
proportional to the number $N$ of points and to the square of the
the amplitude $y_m$ of a given periodic component. Note
that Eq.~(\ref{Eq:PNapprox}) holds approximately for all types of
noises, not only for Gaussian noise. Numerical verifications for
the case of a single noisy sine signal, i.e. $M=1$, have been
presented in Ref.~\cite{Horne1986}.

A severe drawback when using Eq.~(\ref{lombdefspec}) directly is
computational: the time needed to calculate a Lomb periodogram
typically grows as $N^2$ compared to $N \ln N$ for the FFT. To fix
this, a fast computation approach of the Lomb periodogram has been
proposed \cite{Press1989,Press1996}. The method uses the FFT, but
it is not a FFT periodogram of the data. In this method,
$P_N(\omega)$ can be calculated with any desired precision and the
time needed is only of order $N \ln N$. This is the method used in
the present work.

However, the presence of a peak in the Lomb periodogram does not
guarantee the presence of a genuine periodic signal as random
fluctuations can create peaks by chance alone. What is needed is a
quantitative assessment of the probability that peaks can be
generated by random fluctuations. The statistical significance of
a given peak is the probability that it does not stem from noise.
In this goal, previous works \cite{Press1996} have
defined a null hypothesis, which assumes
that a given peak observed in the Lomb periodogram stems from
independent Gaussian noise. Here, we shall develop a hierarchy
of more sophisticated null hypothesis.

If the signal $y(t_i)$ is a pure noise, namely the noise is
independently normally distributed then, at any particular
${\omega}$, it can be shown that the normalized periodogram
$P_N\left(\omega\right)$ has an exponential probability
distribution with unit mean \cite{Scarg1982}. In other words, the
probability that the height $z = P_N\left(\omega\right)$ of the Lomb
periodogram at any given frequency is found
between some positive
$z$ and $z+{d}z$ is $e^{-z}{d}z$ \cite{Horne1986}, independently
of the amplitude of the Gaussian noise. This independence with respect
to the noise amplitude stems from the normalization explicited in
equations (\ref{lombdefspec},\ref{Eq:PNapprox}).
If we scan some
$M$ independent frequencies $\omega_1, \omega_2, ..., \omega_M$, the
probability that none of these
$M$ frequencies give a corresponding value of the Lomb periodogram
$P_N(\omega_1), P_N(\omega_2), ..., P_N(\omega_M)$
larger than $z$ is $\left( {1 - {\mathop{\rm e}\nolimits} ^{ - z}
} \right)^M $. Therefore, the probability for any Lomb peak to exceed $z$ is
\begin{equation}
Pr \left(  { > z} \right) = 1 -{\left( {1 - e^{ - z} }
\right)}^M~, \label{EqNumRecipe1}
\end{equation}
which defines the false-alarm probability of the null
hypothesis. The significance level of any peak
$P_N\left(\omega\right)$ is thus $1- Pr \left( { > z} \right)$. A
small value for the false-alarm probability indicates a highly
significant periodic signal. The interesting region is where the
false-alarm probability is small such that equation
(\ref{EqNumRecipe1}) can be expanded to give \begin{equation} Pr
\left(  { > z} \right) \approx M e^{ - z}~. \label{EqNumRecipe2}
\end{equation}
In order to estimate the false-alarm probability, extensive Monte
Carlo simulations have been carried out to determine the values of
$M$ in various cases \cite{Horne1986}. In those simulations
\cite{Horne1986}, all data sets were Gaussian noise with different
spacings in time: (1) evenly spaced data in time; (2) time
increments drawn as uniformly distributed random numbers in (0,1);
and (3) increments clumped in groups. These works \cite{Scarg1982,
Horne1986} thus aimed at testing the impact of different
non-uniform sampling rates, rather than the effect of non-Gaussian
non-white noise as studied here.

Our interest in this problem is motivated by the problem of
detecting so-called log-periodic oscillations in self-similar
systems, which are the signatures of the symmetry of discrete
scale invariance (DSI) \cite{DSISor}. The hallmark of
self-similarity is the presence of a power law followed by an
observable $F(x)$ as a function of the scale $x$ of observation or
of the distance $x$ to the critical point. The hallmark of DSI is
the existence of oscillations that are periodic in the logarithm
of $x$, hence the name ``log-periodic''. As a complement to
fitting an observable with a power law decorated by log-periodic
oscillations, a non-parametric detection of log-periodicity is of
great interest to test for the presence of DSI and was introduced
in \cite{johsor1,johsor2,johsor3} based on the Lomb periodogram.

Such a method indeed requires the use of the Lomb periodogram analysis
because, by construction, the sampling is non-uniform since the
analysis must be performed in the variable $\ln x$. In addition,
only a few log-periodic oscillations are usually available, since
each fully completed oscillation corresponds to a multiplication
of the observable by a preferred scaling ratio $\lambda>1$. Hence,
$n$ oscillations correspond to spanning the interval $[x_0, x_0
\lambda^n]$, where $x_0$ is some constant. Thus many decades of
data are required to observe only a few oscillations. In order to
perform the Lomb periodogram of an observable exhibiting scale
invariance, one can either subtract a pure power law fit from the
data and analyze the residue with the Lomb periodogram method in
terms of the variable $\ln x$ \cite{Huang2000a,Huang2000b} or
define a local exponent by constructing a local logarithmic
derivative $d\ln F/d\ln x$. Then, the potential oscillatory
structure of the fluctuations of this local exponent as a function
of $\ln x$ can be analyzed \cite{Sorne1996}.

The quantification of the false-alarm probability given by
(\ref{EqNumRecipe1}) and (\ref{EqNumRecipe2}) is based on the
assumption that the noise is Gaussian and uncorrelated, as
mentioned above. In practice, this is almost always wrong. The
present work investigates how the false-alarm probability $Pr
\left(  { > z} \right)$ (and therefore the confidence level $1- Pr
\left( { > z} \right)$) changes from the predictions
(\ref{EqNumRecipe1}) and (\ref{EqNumRecipe2}) in the presence of
non-Gaussian noise and of dependence between successive
innovations (increments) of the time series. If one knows some of the
characteristics of the noise distribution and of its dependence
structure, a more refined null hypothesis asks the following
question: what is the false-alarm probability of the highest peak
of a Lomb periodogram performed on a pure noise with the same
distribution and same dependence structure?

Two recent papers \cite{Huang2000a,Huang2000b} have already
partially addressed this question. Ref.~\cite{Huang2000a}
identified a novel source for periodicity relying solely on the
manipulation of data: (1) approximate uniform sampling and (2) use
of cumulative or integrated quantities which reddens the noise (increases
the power spectrum at low frequencies)
and, in a finite sample, creates a maximum in the spectrum leading
to a most probable frequency. It was found that (i) the
frequencies of the oscillations have a power spectrum given
approximately by $\omega^{-2}$; (ii) the most probable frequency
scales as the inverse of the interval size; (iii) the amplitude of
the periodic oscillations scales as the inverse square root of the
number of data points (central limit theorem). These properties
were studied in the context of the search for log-periodicity in
earthquake aftershock time series, where the natural variable is
$\ln t$. Ref.~\cite{Huang2000b} presented a battery of synthetic
tests performed to quantify the statistical significance of the
existence of log-periodic oscillations in cumulative seismic data
up to major earthquakes. By defining synthetic tests that are as
much as possible identical to the real time series except for the
property to be tested, namely log-periodicity,
Ref.~\cite{Huang2000b} found that log-periodic oscillations with
frequency and regularity similar to those of the real seismic data
are likely to be generated by the interplay of the low pass
filtering step due to the construction of cumulative functions
together with an approximate power law acceleration.

As a prerequisite of the analysis of log-periodicity in many
systems (growth processes, rupture, earthquakes, turbulence,
finance, etc), it is necessary to extend the previous
investigation on the Lomb analysis to non-Gaussian noise with the
possible addition of correlated noise. In the sequel, we examine
the following three cases: 1) noises with heavy tails but no
correlations, 2) Gaussian noises with correlations, and 3) noises
with both non-Gaussian heavy tails and correlations. We perform
extensive numerical simulations to determine the effects of either
heavy-tailness or correlations on the significance level of the
peaks.

Our presentation is organized as follows. We investigate the effect of
heavy-tailness in noises on the significance level of the signals in
Sec.~\ref{sec:fatness}. We analyze two types of noises, the first
class characterized by L\'{e}vy stable distributions and the
second one distributed according to general power law
distributions. In Sec.~\ref{sec:correlation}, we consider
fractal-Gaussian noises with long-range corrections (both
persistent and anti-persistent). The interplay between
heavy-tailness and correlations in noises is studied in
Sec.~\ref{sec:interplay}. Together with the heavy-tailed noise
distributions, we consider both short-range exponentially decaying
correlations as well as very long-range correlation such as in
fractional Gaussian motions. We consider the joint distribution of
the frequency and height of the highest peak of the Lomb power
spectrum in Sec.~\ref{sec:joint}. Application of the numerical
results to real systems are discussed in Sec.~\ref{sec:appl}. We
summarize our results in Sec.~\ref{sec:conclusion}.

\section{Effect of heavy-tailness}
\label{sec:fatness}

\subsection{L\'{e}vy stable noise}
\label{subsec:LSN}

We consider an uncorrelated L\'evy noise \cite{Levy1937}, in which
each innovation (increment) is independently drawn from the same L\'evy
distribution. There is no closed analytical expression for the
probability function of the stable distributions with a few
exceptions, e.g., Cauchy distribution ($\alpha=1$) and Gaussian
distribution ($\alpha=2$) \cite{Paint1995}. Stable distributions
are given in terms of their characteristic function
\cite{Samor1994}.
We denote $X \sim B^\alpha (\lambda,\mu,\sigma)$ or $B^\alpha
(x;\lambda,\mu,\sigma)$ such a L\'{e}vy stable distribution.
The symbol $\sim$ means that the random variable $X$ is drawn from the process
$B^\alpha (\lambda,\mu,\sigma)$. This process
is specified by four parameters: $\alpha$ is the stable
characteristic exponent $(0<\alpha \leq 2)$, $\lambda$ is the
symmetry parameter $(-1 \leq\lambda\leq 1)$, $\sigma$ is the scale
factor $(\sigma>0)$, and $\mu$ is the location parameter. When
$\alpha = 2$, the distribution is Gaussian with mean $\mu$ and
variance $2\sigma^2$, and $\lambda$ has no effect. The tail
probability distribution in the non-Gaussian cases $\alpha<2$
is known asymptotically as \cite{Samor1994,Harma2001}:
\begin{equation}
   {Pr}(B^\alpha > x) \simeq \sigma^{\alpha} \frac{1+\lambda}{2}
C_{\alpha} x^{-\alpha},
   \label{eq:LevyTD}
\end{equation}
where $C_{\alpha} = \frac{1-\alpha}{\Gamma(2-\alpha) \cos(\pi
\alpha/2)}$, if $\alpha \neq 1$. Hence, stable distributions have
heavy tails: the lower the value $\alpha$ is, the more extended the tails
become and the greater is the probability of extreme events. In
the case of $\lambda=0$, we have the expression of symmetric
L\'{e}vy stable distributions \cite{Felle1971,Mante1994}:
\begin{equation}
f(x) = \frac{1}{\pi} \int_0^\infty \exp(-\sigma q^{\alpha})
\cos(qx) dq.
\end{equation}

\begin{figure}
\begin{center}
\epsfig{file=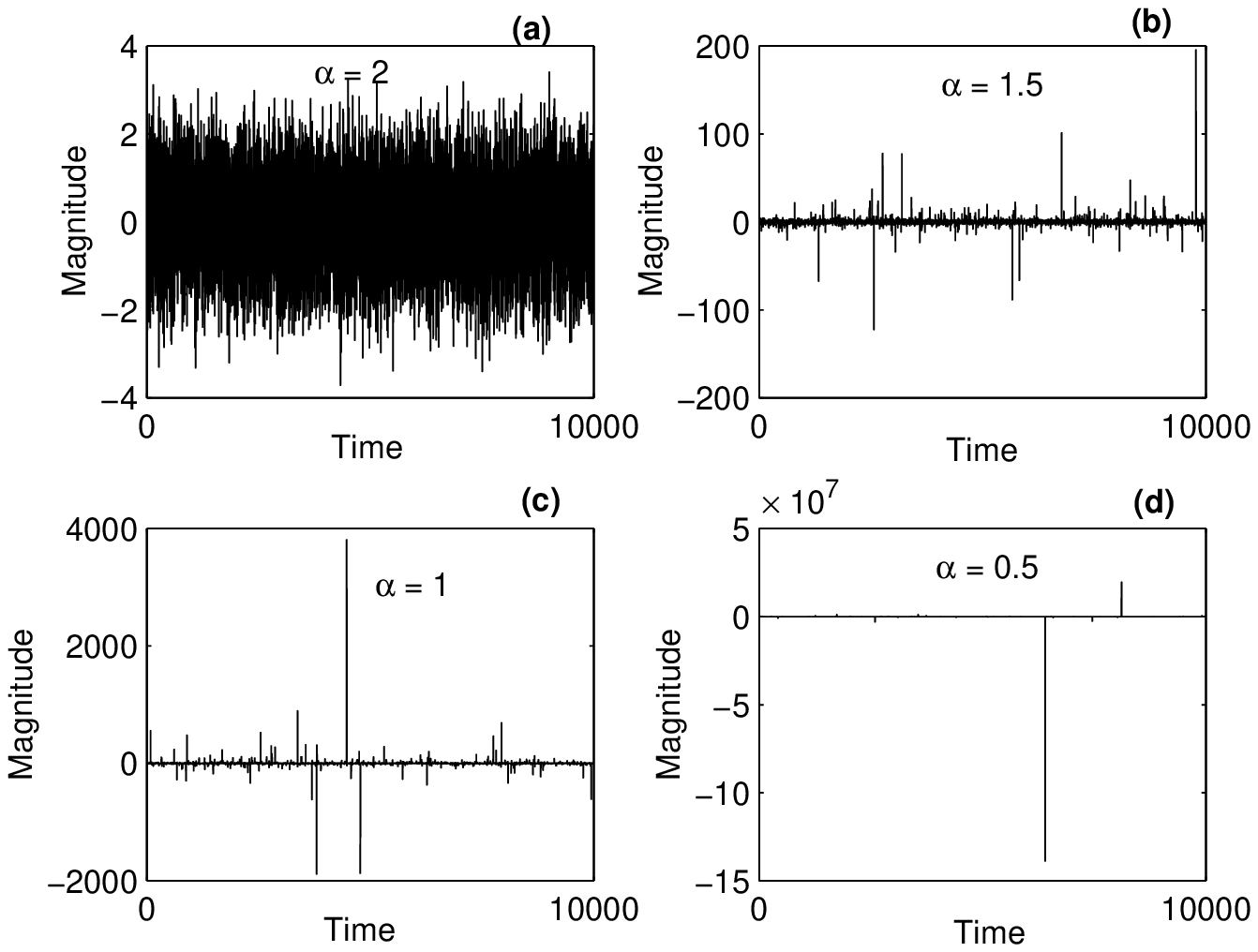,width=13cm, height=9cm}
\end{center}
\caption{\small{Surrogate L\'{e}vy stable noises each with 10,000
data points for: (a) $\alpha = 2$; (b) $\alpha = 1.5$; (c) $\alpha
= 1$; and (d) $\alpha = 0.5$. Here, $\lambda=0$, $\mu=0$ and
$\sigma=1/\sqrt{2}$. Note the difference in the vertical scales of
the four panels.}} \label{Fig:NoiseiLn}
\end{figure}

Our generation algorithm is based on the so-called CMS method
\cite{Chamb1976}, which is applied in such a way that $x$ will
have the characteristic function $\Phi(\theta)$ of the L\'evy stable 
distribution
\cite{Samor1994}. With this parameterization, the stable
distribution $B^\alpha (x;\lambda,\mu,\sigma)$ equals $B^\alpha
(\frac{x-\mu}{\sigma};\lambda,0,1)$ \cite{McCul1996}. We used the
program provided by McCulloch \footnote{Available at
http://economics.sbs.ohio-state.edu/jhm/jhm.html}. In our
simulations, we fixed $\lambda=0$, $\mu=0$ and
$\sigma=\frac{1}{\sqrt{2}}$, and changed $\alpha$ from 0.1 up to 2
with step 0.1 so that we can compare our results with those of
Gaussian white noise. Four typical noises with 10,000 data points
are illustrated in Fig.~\ref{Fig:NoiseiLn}: (a) $\alpha = 2$, (b)
$\alpha = 1.5$, (c) $\alpha = 1$, and (d) $\alpha = 0.5$. For each
$\alpha$, we generated 50,000 data sets each with 100 data points.
The unevenly spaced sampling of times which is adopted throughout
this paper is generated as follows: one first generates 100
uniformly distributed random numbers between 0 and 1, and then
obtains the cumulative sums that are used as the sampling times
$t_1,\cdots,t_{100}$. By construction, the sampling intervals
$\triangle t=t_{i+1}-t_i$ are uniformly distributed. We performed
the Lomb analysis on each data set.

\begin{figure}
\begin{center}
\epsfig{file=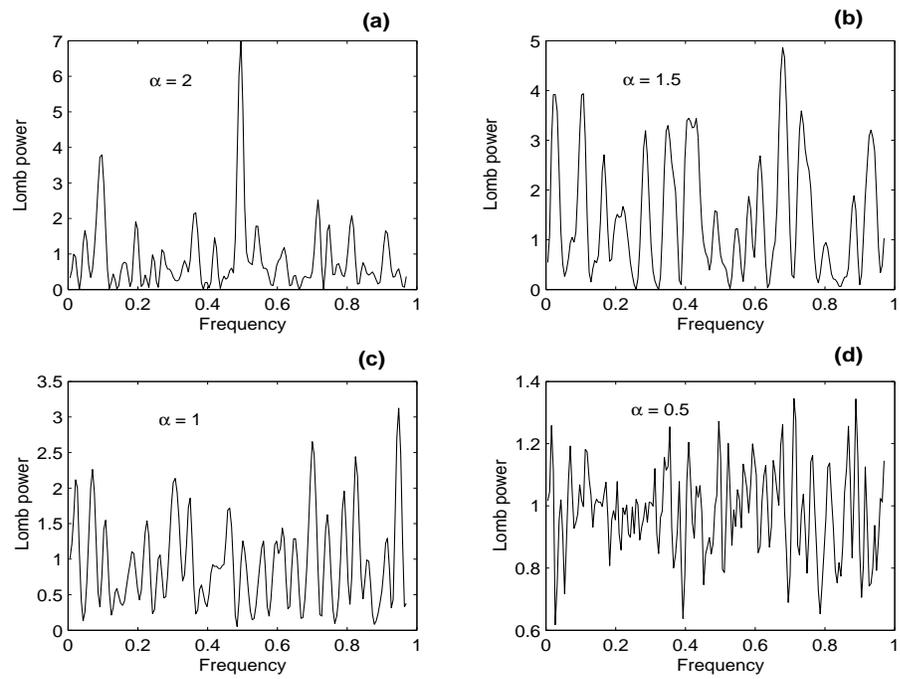,width=13cm, height=9cm}
\end{center}
\caption{\small{The four Lomb periodograms corresponding
to the four cases shown in Fig.~\ref{Fig:NoiseiLn} but with time series
of only $100$ points.}}
\label{FigiLnLomb}
\end{figure}

\begin{figure}
\begin{center}
\epsfig{file=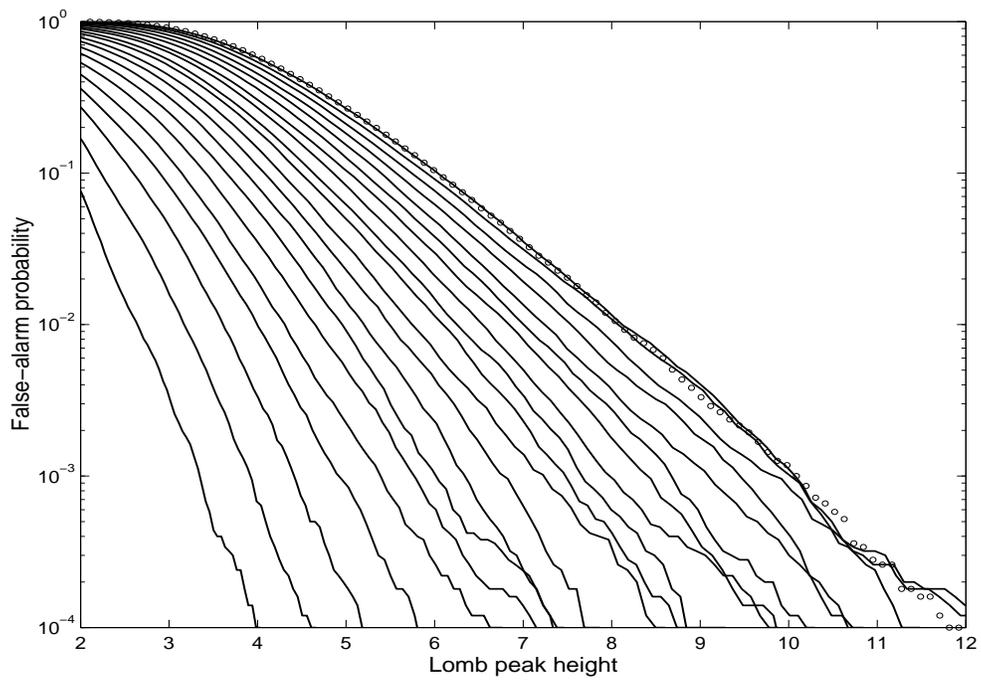,width=13cm, height=9cm}
\end{center}
\caption{\small{False-alarm probability of the surrogate L\'{e}vy
noises for $\alpha$ ranging from 0.1 to 2 with step 0.1 from left to right. 
For a given Lomb peak height, the false-alarm probability increases with
$\alpha$.}} \label{LevySigLevel}
\end{figure}

Fig.~\ref{FigiLnLomb} shows the four Lomb periodograms corresponding
to the four time series shown in Fig.~\ref{Fig:NoiseiLn}. It is
visually suggesting that the spontaneous emergence of peaks above
the background becomes less and less probable for smaller $\alpha$.
This is confirmed by extensive statistical tests synthetized in
Fig.~\ref{LevySigLevel} which plots the false-alarm probability as a
function of the Lomb peak for different exponents $0.1 \leq \alpha
\leq 2$. For reference and comparison, the Gaussian case is
represented by the open circles. Within statistical fluctuations,
it is identical to the last curve to the right corresponding to
$\alpha = 2$, as it should. Remarkably, for a given Lomb peak
height, the false-alarm probability increases with $\alpha$.
Hence, a L\'{e}vy noise with heavier tail (smaller exponent
$\alpha$) results in a higher significance for a given Lomb peak
height. In other words, it is more difficult to generate a large
peak in the Lomb periodogram analysis when the noise has more extended
tail. Intuitively, the larger amplitudes of the noise fluctuations
increase the effective randomness of the noise and destroy
possible oscillations constructed by random occurrences. In other
words, Fig.~\ref{Fig:NoiseiLn} suggests that, the smaller $\alpha$
is, the smaller is the number of points dominating the overall
signal: for instance the last panel for $\alpha =0.5$ shows that
only a few very large peaks dwarf the rest of the population of
noise innovations (increments). Since there is very little structure, there is
little chance for a periodic component to appear, hence the small
false alarm probability.

The semi-logarithmic representation of Fig.~\ref{LevySigLevel}
suggests that the false-alarm probability is approximately an
exponentially decaying function of the Lomb peak, similarly to the
Gaussian case described by Eq.~(\ref{EqNumRecipe2}). This suggests
to fit the dependence of the false-alarm probability for various
$\alpha$ as an exponential of the Lomb peak height
\begin{equation}
   Pr_{\alpha}(z) = M(\alpha) e^{-k(\alpha) z}~,
   \label{EqLevyPr}
\end{equation}
with an effective number $M(\alpha)$ of independent frequencies
and a decay rate $k(\alpha)$, which are expected to be
respectively increasing and decreasing functions of $\alpha$.

In order to obtain more reliable estimations of $M(\alpha)$ and
$k(\alpha)$, we excluded the largest $z$'s from the fit with
Eq.~(\ref{EqLevyPr}), because they are spoiled by large
fluctuations. The scaling range was fixed in the interval
$Pr_\alpha(z) \in [0.001,0.05]$. The upper bound $0.05$ ensures
the validity of the approximation (\ref{EqNumRecipe2}). The lower
bound $0.001$ is a good compromise to have both a large range for
the fit and to exclude the largest fluctuations occurring for the
smallest $Pr_\alpha(z)$. The error bars for $k(\alpha)$ are
estimated by evaluating the standard deviation over 12
subintervals constituted of the ten subintervals resulting from
subdividing the range $[0.001,0.05]$ equally in ten parts and
adding two adjoining subintervals of the same size, one below the
lower bound and one above the upper bound. The results are listed
in Table \ref{Tb:Stabfit}. We find $M(2)=104$ and
$k(2)=1.14\pm0.19$, which are consistent with the results $M=119$
and $k=1$ known for independent Gaussian noise \cite{Horne1986}.
Table \ref{Tb:Stabfit} shows a systematic decrease of $k(\alpha)$
as $\alpha$ increases, which is capturing the observation that the
false-alarm  probability is increasing with $\alpha$. We find
however that the determination of the effective number $M(\alpha)$
of independent frequencies is not reliable as its determination is
very sensitive to the value $k(\alpha)$. Therefore, we cannot
conclude about a specific dependence of $M(\alpha)$ as a function
of $\alpha$. This limitation is relatively minor as this
pre-exponential factor $M(\alpha)$ has a weak impact on the
false-alarm probability (a factor two or three at most).

\begin{table}
\begin{center}
\begin{tabular}{|c|c|c|c|c|c|c|c|}
    \hline
   $\alpha$ &0.1          & 0.2         & 0.3         & 0.4         &
0.5\\ \hline
   $k$
&3.06$\pm$0.72&2.89$\pm$0.74&2.69$\pm$0.40&2.42$\pm$0.41&2.12$\pm$0.25
\\ \hline
   $M$      &32           &     90      &    154      &    151      &
90\\ \hline
   $\alpha$ &0.6          &     0.7     &    0.8      &    0.9      &
1.0\\ \hline
   $k$
&1.98$\pm$0.27&2.04$\pm$0.29&1.89$\pm$0.36&1.73$\pm$0.24&1.59$\pm$0.18
\\ \hline
   $M$      &97           &    224      &    185      &    133      &
92\\ \hline
   $\alpha$ &1.1          &    1.2      &    1.3      &    1.4      &
1.5\\ \hline
   $k$
&1.59$\pm$0.24&1.50$\pm$0.33&1.49$\pm$0.15&1.39$\pm$0.17&1.33$\pm$0.23
\\ \hline
   $M$      &150          &    121      &    166      &    131      &
112\\ \hline
   $\alpha$ &1.6          &    1.7      &    1.8      &    1.9      &
2.0\\ \hline
   $k$
&1.27$\pm$0.11&1.19$\pm$0.15&1.16$\pm$0.12&1.12$\pm$0.20&1.14$\pm$0.19
\\ \hline
   $M$      &101          &    78       &    85       &     80      &
104\\ \hline
\end{tabular}
\caption{List of the parameters $k(\alpha)$ and $M(\alpha)$ of the fit of the
false-alarm probability $Pr_{\alpha}(z)$ by expression (\ref{EqLevyPr}).}
\label{Tb:Stabfit}
\end{center}
\end{table}

\subsection{Symmetrical power-law noises}

We now consider noises distributed with density distribution
functions decaying as pure power laws according to
\begin{equation}
   f(y) =
   \left\{ {\begin{array}{*{20}c}
     \frac{\kappa}{2 |y|^{\kappa+1}}, & |y|>1~,  \\
     0, & |y| \leq 1~. \\
   \end{array} } \right.
   \label{powernsksa}
\end{equation}
These distributions complement the results obtained using the
L\'evy stable distributions of the previous section by 1) testing
for the role of a strong difference in the bulk part of the
distributions and 2) allowing for the exponent $\kappa$ to be
larger than $2$.

We first generate uniformly distributed random numbers $x$ in $(0,1)$,
and then substitute them into
\begin{equation}
y = \left\{ {\begin{array}{*{20}c}
     -{(2x)}^{-1/\kappa}, & 0 < x \leq 0.5,  \\
     {[2(1-x)]}^{-1/\kappa}, & 0.5 < x < 1. \\
\end{array} } \right.
\end{equation}

\begin{figure}
\begin{center}
\epsfig{file=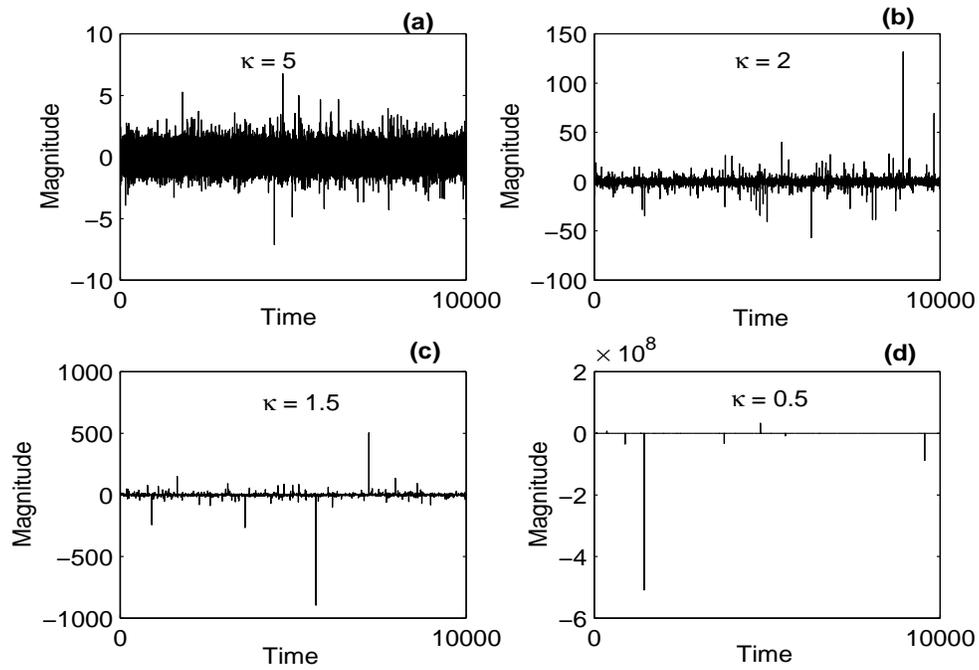,width=13cm, height=9cm}
\end{center}
\caption{\small{Surrogate symmetrical power-law noises each with
10,000 data points for: (a) $\kappa = 5$; (b) $\kappa = 2$; (c)
$\kappa = 1.5$; and (d) $\kappa = 0.5$. Note the difference in the
vertical scales of the four panels.}} \label{Fig:NoiseTPT}
\end{figure}

\begin{figure}
\begin{center}
\epsfig{file=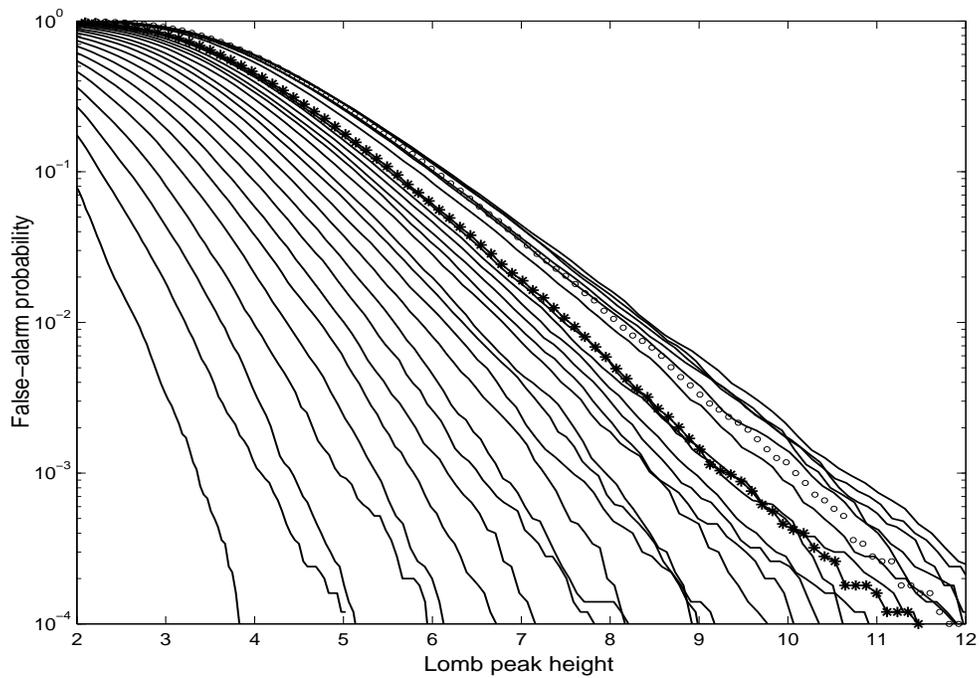,width=13cm, height=9cm}
\end{center}
\caption{\small{False-alarm probability of the surrogate power-law
tailed noise for $\kappa$ ranging from from 0.1 to 2 with spacing 0.1, 
from 2.5 to 4 with spacing 0.5, and for 5 and 6, from left to right. 
For a given Lomb
peak height, the false-alarm probability increases with $\kappa$.
The line marked with stars is for $\kappa=2$ while the circles
correspond to the Gaussian case.}} \label{PowTailSigLevel}
\end{figure}

In the simulations, we consider values of $\kappa$ from $0.1$ to
$2$ with spacing $0.1$, from $2.5$ to $5$ with spacing $0.5$, and
the values $\kappa = 6$ and $\kappa=7$. Four typical noises with
10,000 data points are illustrated in Fig.~\ref{Fig:NoiseTPT}: (a)
$\kappa = 5$, (b) $\kappa = 2$, (c) $\kappa = 1.5$, and (d)
$\kappa = 0.5$. For each $\kappa$, we generated 50,000 data sets,
each with 100 data points, as in Sec.~\ref{subsec:LSN}. Figure
\ref{PowTailSigLevel} plots the false-alarm probability as a
function of the Lomb peak for different exponents $0.1 \leq \kappa
\leq 7$. For reference and comparison, the Gaussian case is
represented by the open circles. The results are very similar to
those found for the L\'evy noise shown in Fig.~\ref{LevySigLevel}.
As the exponent $\kappa$ increases, the false-alarm probability
increases: the thinner the tail is, the smaller is the confidence
level of a given Lomb peak. Intuitively, the results are as for
the L\'evy noise: largest fluctuations for small $\kappa$
introduce a larger effective randomness making less probable the
occurrence of oscillations. Similarly to the argument proposed in
the previous section, Fig.~\ref{Fig:NoiseTPT} suggests that, the
smaller $\kappa$ is, the smaller is the number of points
dominating the overall signal. As a consequence, the time series
will have little structure, hence the small false-alarm
probability.

In contrast with the L\'evy noise for which the false-alarm curves
accumulate at the Gaussian curve for $\alpha=2$, the false-alarm
curves for the power law noises continue to translate to the right
as $\kappa$ increases. The reason is that $\kappa=2$ for a power
law distribution is not the same as $\alpha=2$ for a L\'evy stable
distribution: the former case corresponds to a power law
distribution $\sim |y|^{-3}$ while the latter corresponds to the
Gaussian distribution. Interestingly, for $\kappa>3$, we observe
that the false-alarm probabilities are larger than for the
Gaussian case. In this sense, uncorrelated noises with power law
distributions with tail exponent $\kappa>3$ are ``less random''
than the Gaussian case as oscillations caused by random
occurrences are more frequent.

\begin{figure}
\begin{center}
\epsfig{file=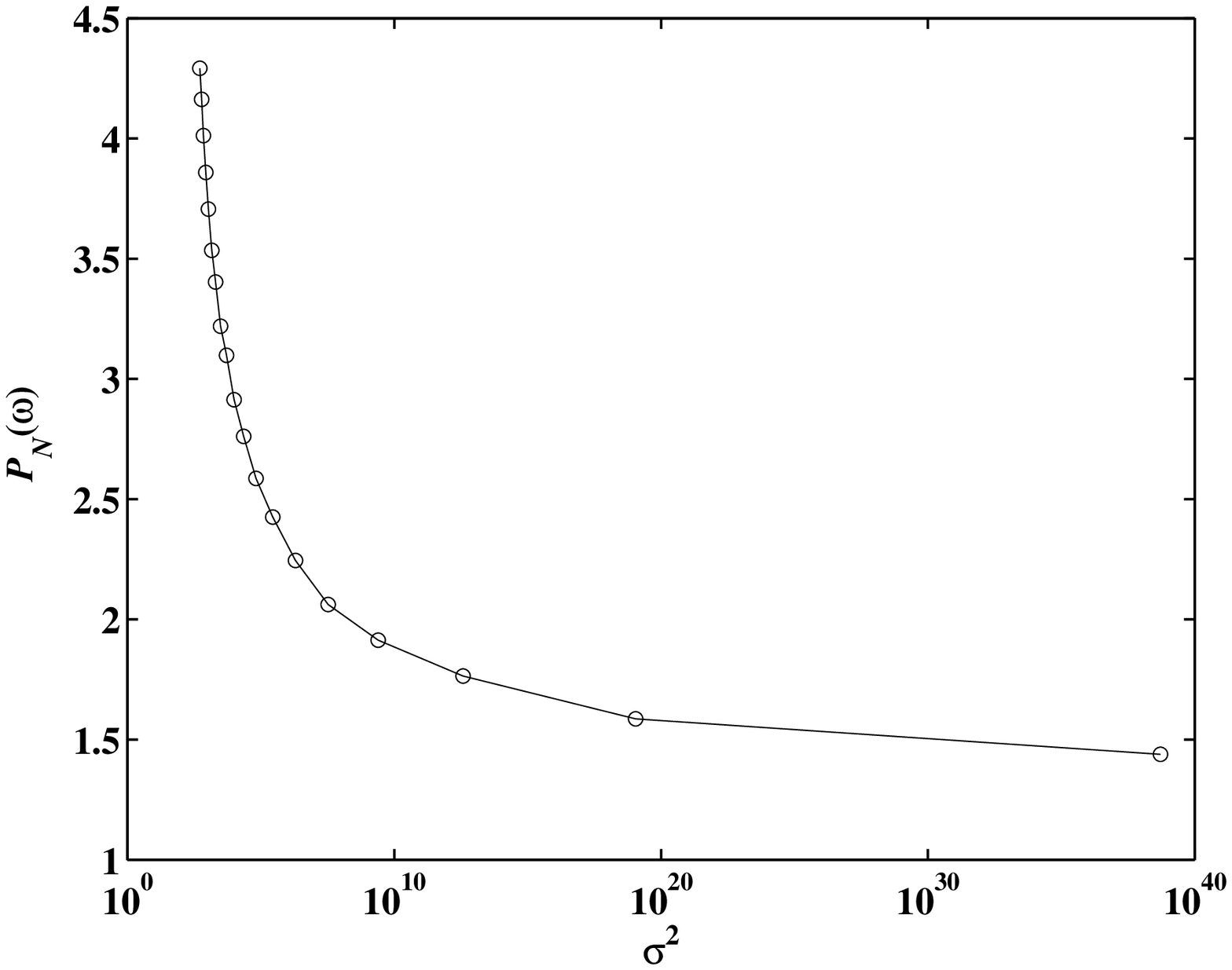,width=13cm, height=9cm}
\end{center}
\caption{\small{Dependence of the maximal Lomb peak $P_N(\omega)$
averaged over 50,000 realizations of 100 points as a function of
$\tilde{\sigma}^2$ given by expression (\ref{empivari}). }}
\label{FigPNsigma}
\end{figure}

\begin{figure}
\begin{center}
\epsfig{file=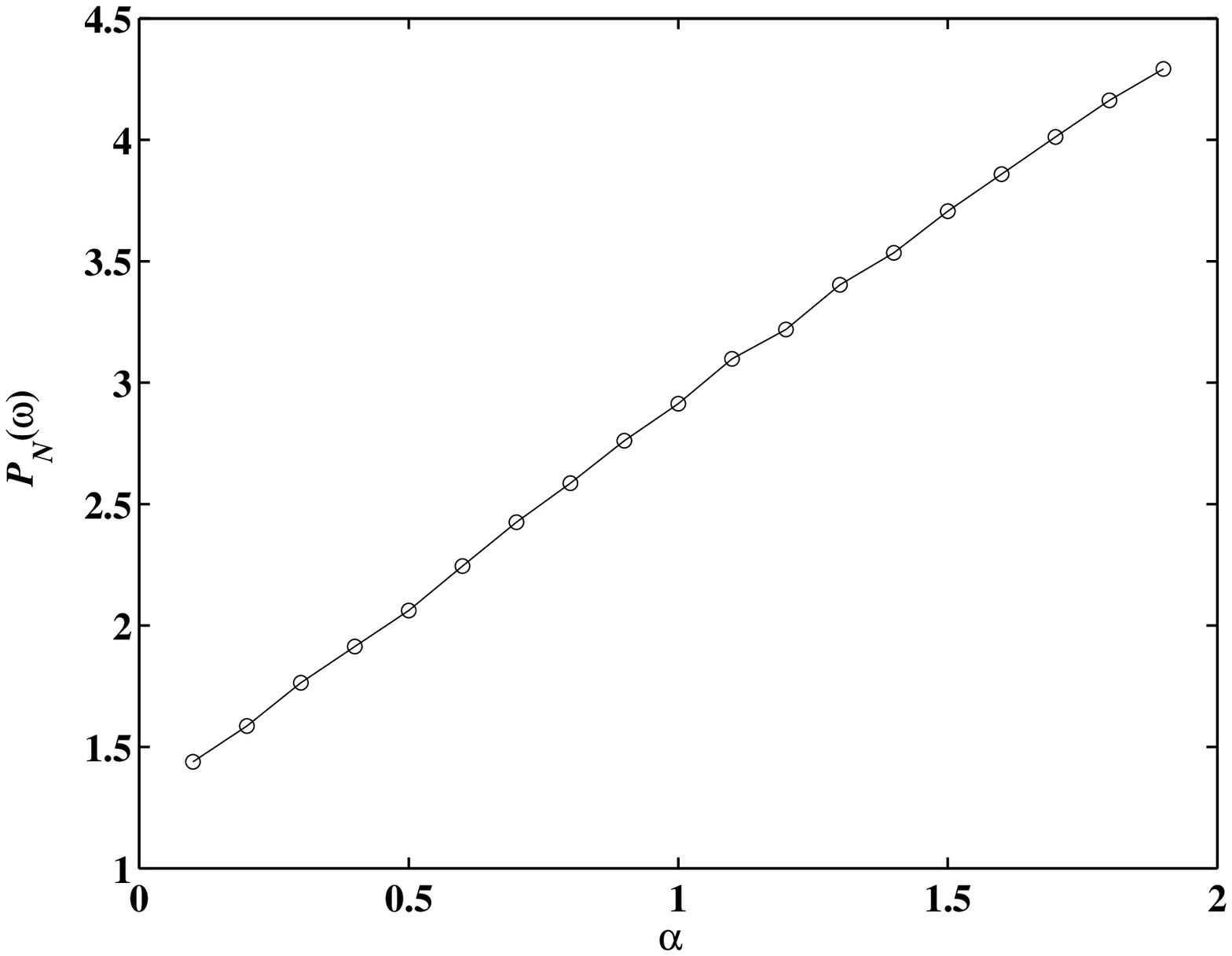,width=13cm, height=9cm}
\end{center}
\caption{\small{Dependence of $P_N(\omega)$ as a function of the
power law exponent $\alpha$.}} \label{FigalphaPN}
\end{figure}

The reason why a more extended tail leads on average
to smaller Lomb peaks, such that a given peak height
has stronger false-alarm probability (smaller statistical
significance) for the Gaussian noise than for L\'evy noise,
can be easily understood from the following argument.
Qualitatively, a smaller exponent leads to larger fluctuations
which introduce a larger effective randomness making less probable the
occurrence of oscillations. Quantitatively, this ``larger randomness''
can be measured by the sample-size-dependence of the empirical
variance $\sigma^2$ entering the height of the Lomb peak according to
expression (\ref{Eq:PNapprox}).
Let us present our
argument for the power laws (\ref{powernsksa}) for which the
explicit dependence on the exponent is simpler. Here, we use the
same notation $\alpha = \kappa$ to stress that the argument
applies in general to distributions with power law tails.
Consider the case $\alpha < 2$ for power laws for which the
variance does not exist from a mathematical point of view.
Notwithstanding this fact, one can always calculate an empirical
variance from each specific time series, as
\begin{equation}
\tilde{\sigma}^2 = \frac{1}{N-1}\sum_{i=1}^{N} \left( x_i -
\langle{x_i}\rangle \right)^2,
\end{equation}
where $x_i$ is the $i$th point of the synthesized L\'evy stable
noise. The notation $\langle{x_i}\rangle$ refers also to the
empirical estimation of an average from the empirical time series.
The fact that the variance (and the mean for $\alpha \leq 1$) does
not exist is reflected in the fact that the estimated variance
grows with data size $N$ and exhibits large fluctuations.
$\tilde{\sigma}^2$ is scaling as
\be
\tilde{\sigma}^2
   \propto  N \alpha \int^{x_{\rm max}}_{1} dx~{x^2 \over x^{1+\alpha}}
   \propto {N \alpha \over 2-\alpha} \left( N^{2-\alpha \over \alpha} 
-1 \right)~,
   \label{empivari}
\ee where we have estimated the typical largest value $x_{\rm
max}$ by the standard condition \be N \alpha \int_{x_{\rm
max}}^\infty dx~x^{-1-\alpha}  = 1~, \label{ngnnla}
\ee
leading
to the estimation $x_{\max} = N^{1/\alpha}$. Expression
(\ref{ngnnla}) expresses the fact that the product of the
probability to be of the order of or larger than $x_{\rm max}$ by
the total number of points is of the order $1$, which defines
$x_{\rm max}$. Figure \ref{FigPNsigma} shows the maximal Lomb peak
$P_N(\omega)$ averaged over 50,000 realizations of 100 points as a
function of $\tilde{\sigma}^2$ given by expression
(\ref{empivari}). Here $\tilde{\sigma}^2$ is varied by changing
the power law exponent from $\alpha=0.1$ to $1.9$. Increasing
$\alpha$ decreases $\tilde{\sigma}^2$ and thus corresponds to
reading figure \ref{FigPNsigma} from right to left. Figure
\ref{FigalphaPN} shows directly the dependence of $P_N(\omega)$ as
a function of $0.1 \leq \alpha \leq 1.9$ in the same range. This
rationalizes our empirical finding that the Lomb peak
$P_N(\omega)$ and as well as the maximal peak increases with
$\alpha$. This is in line with the results presented in
Figs.~\ref{LevySigLevel} and \ref{PowTailSigLevel}. We will
observe a similar effect in GARCH(1,1) noise discussed in
Sec.\ref{subsec:garch}.

\section{Effect of correlations}
\label{sec:correlation}

\subsection{Fractional Gaussian noises}
\label{subsec:fGn}

We now study the effect of correlations included in fractional
Gaussian noise (fGn) $X_H(t)$ of parameter $H$. The cumulative sum
$B_H(t)$ of fGn defines the fractional Brownian motion (fBm) which
was introduced by Mandelbrot and Van Ness \cite{Mande1968} as an
extension to the usual Brownian motion. A process $B_H(t)$ is a
fBm if it is a Gaussian random process which has stationary and
self-similar increments,
\begin{equation}
   X_H(t)=B_H(t+1)-B_H(t)~,  \label{ngnlla}
\end{equation}
which form a fractional Gaussian noise (fGn). The value $H =
\frac{1}{2}$ yields the ordinary Brownian motion, which is known
to be non-persistent with absence of correlations of the
increments $X_{1/2}(t)$. For $0< H < \frac{1}{2}$, the fBm is
antipersistent, while for $\frac{1}{2} < H < 1$, the fBm is
persistent \cite{Mande1982}.

There are numerous synthesis methods proposed to generate fBm, and
hence fGn. They include the random midpoint displacement method
\cite{Mande1982,LauWC1995,Norro1999}, the Cholesky decomposition
method of covariance matrix of vector increments using the
Durbin-Levinson algorithm \cite{Lunda1986,Brock1991,Taqqu1995},
the inverse Fourier transform method based on power spectrum
\cite{Barns1988,Fland1989}, the fast Fourier transform method
matching covariance structure instead of frequency spectrum
\cite{Dietr1997,Crous1999}, more recently the wavelet transform
method \cite{Fland1992,Stoks1994,Heneg1996,HuZhu2000} and others
\cite{Jenna1996,Crous1999}.

We adopted the Cholesky-Levinson factorization method\footnote{We
used the MATLAB script named fbmlevinson.m included in FracLab
available at
http://www-rocq.inria.fr/fractales/Software/FRACLAB/.}, which is
one of the best methods among those used to synthesize Gaussian
time series \cite{Jenna1996}. This algorithm for generating fBm
performs a linear transformation of i.i.d. Gaussian random
variables in three steps: generation of independent Gaussian
random numbers, multiplication of these random numbers by weights
that are recursively determined by the covariance structure, and
construction of their running sum \cite{Brock1991}. The fGn are
the innovations (increments) of fBm. In the sequel, we analyze
the statistical significance of the Lomb periodogram of fGn.

\subsection{Numerical results}

In our simulations, the Hurst index ranges from 0.1 to 0.9 with
spacing 0.1. For each $H$, we generated 50,000 fGn time series.
The length of each time series is again 100 data points. Figure
\ref{Fig:NoisefGn} shows four fractional Gaussian noises: (a)
$H=0.2$, (b) $H=0.5$, (c) $H=0.7$, and (d) $H=0.9$. It is visually
apparent that, the larger is $H$, the clearer is the presence of
oscillations within the noise.

We performed the Lomb analysis on each fGn time series and
extracted the largest peaks in the Lomb periodograms. The
dependence of the false-alarm or rejection probability as a
function of the Lomb peak height for different types of noises is
shown in Fig.~\ref{FigLombfGn}. For $H=0.5$, we recover the
exponential decay (\ref{EqNumRecipe2}) represented by the open
circles. As $H$ increases, the false-alarm probability increases
very significantly and also exhibits important deviations from an
exponential fall-off.

For $H<0.5$, the fGn is anti-persistent. Thus, a point with
negative (resp. positive) value in the series is usually followed
by a point with positive (resp. negative). This effect favors a
random oscillatory pattern at the highest possible frequency
corresponding to a period of two points but the number of
independent frequency is thus small. For smaller frequencies, this
antipersistence makes improbable the spontaneous formation of
other oscillatory regularities. In contrast, the fGn with $H>0.5$
is persistent. Thus a point of negative (resp. positive) value in
the series is usually followed by a point of negative (resp.
positive) value. Hence, there are more regularities and structure
than for the Gaussian white noise. Long-range correlations create
strong artifactual periodicity. This recovers while generalizing
to different noise spectra the results of
\cite{Huang2000a,Huang2000b}. A larger $H$ implying a larger
correlation between the successive data points, and hence stronger
regularities, it is natural to obtain a higher Lomb peak. This is
true for the power spectra of the fGns proportional to
$1/f^{\beta}$ with $\beta= 2H-1<2$ which corresponds to long-range
correlated stationary noises analyzed here. This is also true for
noise spectral with exponent $\beta\geq 2$ for which the processes
are not stationary (as in a random walk whose standard deviation
grows as $\sqrt{t}$) with correlation functions blowing up.

\section{Interplay between heavy tail and correlations}
\label{sec:interplay}

\subsection{The GARCH(1,1) process}
\label{subsec:garch}
\subsubsection{Method for generation of GARCH(1,1) residuals}

Autoregressive Conditional Heteroscedasticity (ARCH) process was
introduced by Engle in 1982 \cite{Engle1982} to account for the
so-called heteroscedasticity in economic time series.
Heteroscedasticity stands for the lack of stationary
``volatility'' (which is the term for standard deviation using in
finance), i.e., the presence of periods of time with large
volatilities alternating with periods of small volatility. In an
ARCH$\left(q\right)$ process the volatility at time $t$ is a
function of the observed data at $t-1$, $t-1$, $\cdots$, $t-q$. In
1986, Bollersev \cite{Bolle1986} introduced the Generalized ARCH
or GARCH$\left(p,q\right)$ process, where the volatility at time
$t$ depends on the observed data at $t-1$, $t-1$, $\cdots$, $t-q$,
as well as on volatilities at $t-1$, $t-1$, $\cdots$, $t-p$. Here,
we focus our attention on one of the standard benchmark model of
financial time series, the GARCH$\left(1,1\right)$ process, which
obeys the following evolution equation:
\begin{equation}
x_{t}=\mu +\epsilon _{t}~,
\label{Eqxt}
\end{equation}
\begin{equation}
\epsilon _{t}=\sqrt{h_{t}}z_{t}~,
\label{Eqe}
\end{equation}
\begin{equation}
h_{t}=\alpha +\beta \epsilon _{t-1}^{2}+\gamma h_{t-1}~,
\label{Eqht}
\end{equation}
where $z_t$ is a standardized, independent, identically
distributed (i.i.d.) random variable drawn from some specified
probability distribution:
\begin{equation}
\langle{z_t}\rangle=0  ~~~~{\rm and}~~~~  \langle{z_t}^2\rangle=1~.
\label{Eqz}
\end{equation}

The GARCH models have been shown to capture not only volatility
clustering but also accommodate some of the leptokurtosis (i.e.,
heavy tails) commonly found in stock market and currency exchange rate time. However,
GARCH models with conditionally normal errors generally fail to
sufficiently capture the leptokurtosis evident in asset returns.
The increasing attention focused on distributional properties
(particularly tail heavyness), when estimating exchange rates
models, has led to the widespread adoption of non-Gaussian
conditional error distributions, most commonly the Student-t
\cite{Bolle1987,Johan2000,Wang2001}. The Student-t distribution
models more extended tails than the normal distribution and is
asymptotically a power law with an exponent $\kappa$ equal to the
number of its degrees of freedom. The GARCH model embodies
(exponentially decaying) correlations between successive
amplitudes (volatility) of the noise but assumes zero correlations
between the signs of the successive noise innovations (increments).

Since the probability density function of the Student-t
distributed random variable with $\kappa$ degrees of freedom is
\begin{equation}
t\left( x,\kappa \right) =\frac{\Gamma \left( \frac{\kappa +1}{2}\right) }{
\Gamma \left( \frac{\kappa }{2}\right) \sqrt{\kappa \pi }}\left( 1+\frac{
x^{2}}{\kappa }\right) ^{-\frac{\kappa +1}{2}},
\label{Eqtpdf}
\end{equation}
we have (for $n < \kappa$)
\begin{equation}
\left\langle {x^n } \right\rangle  = \left\{ {\begin{array}{*{20}c}
     \begin{array}{l}
   \kappa ^{\frac{n}{2}} \frac{{\Gamma \left[ {{{\left( {n + 1}
\right)} \mathord{\left/
   {\vphantom {{\left( {n + 1} \right)} 2}} \right.
   \kern-\nulldelimiterspace} 2}} \right]\Gamma \left[ {{{\left(
{\kappa  - n} \right)} \mathord{\left/
   {\vphantom {{\left( {\kappa  - n} \right)} 2}} \right.
   \kern-\nulldelimiterspace} 2}} \right]}}{{\Gamma \left( {{1 \mathord{\left/
   {\vphantom {1 2}} \right.
   \kern-\nulldelimiterspace} 2}} \right)\Gamma \left( {{\kappa  \mathord{\left/
   {\vphantom {\kappa  2}} \right.
   \kern-\nulldelimiterspace} 2}} \right)}} \\
   0 \\
   \end{array} & \begin{array}{l}
   \bmod \left( {\kappa ,2} \right) = 0 \\
    \\
   \bmod \left( {\kappa ,2} \right) = 1 \\
   \end{array}  \\
\end{array}} \right.
\end{equation}
In particular, for $\kappa>2$, we have
$\left\langle{x^2}\right\rangle=\kappa/\left(\kappa-2\right)$.
Hence, to meet Eq.~(\ref{Eqz}),
\begin{equation}
z = \sqrt{\kappa/\left(\kappa-2\right)} \cdot x~,  \label{hgnlnlaa}
\end{equation}
whose probability density function is
\begin{equation}
f \left( z \right) = {\Gamma\left[(\kappa + 1)/2\right] \over
\Gamma\left[\kappa/2\right]~\sqrt{\pi (\kappa -2)}}~
\left(1 + {z^2 \over \kappa -2}\right)^{-{\kappa +1 \over 2}}~.
\label{eqzpdf}
\end{equation}
$z$ tends towards the standardized Gaussian random variable as
$\kappa \to \infty$.

We have generated GARCH(1,1) noise time series as follows: 1) For
a given choice of the parameters and of $\kappa$, generate $t$
random t-distributed random numbers; 2) Obtain $z_1$, $z_2$,
$\ldots$, $z_t$ from (\ref{hgnlnlaa}); 3) Generate iteratively
$\epsilon_1$, $\epsilon_2$, $\ldots$, $\epsilon_t$  which have
zero mean and conditional variance $h_t$, using equations
(\ref{Eqe}) and (\ref{Eqht}); 4) A fixed number of the first data
points were discarded in order to remove any sensitivity on the
initial condition.

\subsubsection{Numerical simulations}
\label{subsubsec:garchNS}

We used the parametric values quoted in \cite{Johan2000}: $\mu =
4.38 \times 10^{-4}$, $\alpha = 2.19 \times 10^{-6}$, $\beta =
0.044$, $\gamma = 0.922$, and $h_0 = 6.45 \times 10^{-5}$. In
Fig.~\ref{Fig:NoiseGarch}, we show four typical simulated
GARCH(1,1) noises: (a) $\kappa = 3$, (b) $\kappa = 6$, (c) $\kappa
= 9$, and (d) $\kappa = 12$. Local clustering of volatility are
clearly visually apparent in the figure.

In order to obtain the significance level of the artificial
GARCH(1,1) noises, we simulated a large number of data sets. The
artificial times were unevenly sampled so that each time followed
the previous time by a random number between 0 and 1 as in
previous simulations. We investigated four types of noises with
$\kappa = 3, 6, 9, 12$, as typically shown in
Fig.~\ref{Fig:NoiseGarch}. For each $\kappa$, we generated 50,000
data sets, each with 100 data points. We then Lomb transformed the
50,000 time series and extracted the maximal Lomb peak height in
each periodogram to form a set \cite{Scarg1982}. The results are
shown in Fig.~\ref{FigLombGarch}. To compare with the independent
Gaussian noise, we also simulated 50,000 Gaussian time series and
plotted it in the same figure as the open circles.

For a given peak height, the false-alarm probability increases
with $\kappa$. This is the effect already documented in
Sec.~\ref{sec:fatness}. The effect is weak because $\kappa \geq 3$
and the strongest dependence of the false-alarm probability on the
power law tail exponent $\kappa$ was shown to occur for smaller
exponents. Note that the false-alarm probability of the
GARCH$\left(1,1 \right)$ process tends to the Gaussian curve
$\kappa$ increases. For the largest $\kappa$ investigated, the
false-alarm probability is slightly above the Gaussian
uncorrelated value but the difference is small.  The effect of
correlation of the variance of the GARCH$\left(1,1 \right)$
process has almost no impact on the false-alarm probability, which
is essentially controlled by the non-Gaussian character of the
distribution function. This is checked by reshuffling the GARCH
noise to eliminate the correlations in volatility. Performing the
same analysis, we find the same relation between Lomb peak height
and false alarm probability within statistical fluctuations. Since
the correlations of the variance are destroyed by reshuffling the
GARCH(1,1) noise while keeping the same heavy tail distribution,
we conclude that GARCH$\left(1,1 \right)$ correlations have
essentially no impact on the false-alarm probability of the
detection of periodic components. The slight difference between
the false-alarm probabilities reported in Fig.~ \ref{FigLombGarch}
and in Fig.~ \ref{PowTailSigLevel} for the uncorrelated power law
noise with the same tail exponent can thus be attributed only to
the difference in the bulk of their distributions.

\subsection{Fractional L\'{e}vy noise}
\subsubsection{Method for fractional L\'{e}vy noise synthesis}

We now combine heavy tails and long-range correlations. For this, we
study the fractional L\'{e}vy motion (fLm) \cite{Taqqu1987,
Kogon1996, Huill1999}, which is analogous to fBm \cite{Mande1968},
with the Gaussian distribution of increments replaced by a L\'evy
stable distribution. This corresponds to a moving average with
infinite memory over a set of independent random variables having
a L\'evy distribution. Let us denote $B_H^\alpha(t)$ for fLm.
Then, we have $B_{1/2}^\alpha(t) = B^\alpha(t)$ and $B_H^{2} (t)=
B_H(t)$. Fractional L\'evy motion is a self-similar process with
self-similar index $2H/\alpha>0$ \cite{Huill1999}:
\begin{equation}
{B_H^\alpha(at)}_{t\in{R}} \sim
{a^{2H/\alpha}B_H^\alpha(t)}_{t\in{R}}
\end{equation}
for $a>0$. The symbol $\sim$ denotes that the two processes have
the same distribution. The heavy-tail behavior of fLm is
\cite{Huill1999}
\begin{equation}
\texttt{Pr}(|B_H^\alpha(t)|>x) \propto x^{-\alpha/2H}.
\end{equation}

In order to generate a fractional noise, there are several methods
and the simplest one is to use a fractional ARIMA
\cite{Taqqu2001}. It is easy to implement the algorithm by
replacing the i.i.d. Gaussian noise in the Durbin-Levinson method
in Sec.~\ref{sec:correlation} with L\'evy noise \cite{Taqqu2001}.
In this section, we applied the wavelet method for synthesizing
fractional L\'evy noise. Accurate synthesis of fBm using wavelet
is based on the wavelet analysis of the second order statistics of
fractional Brownian processes \cite{Fland1992}. Hence, fBm is
generated as a sum of wavelets\cite{Stoks1994,HuZhu2000}.
Similarly, we can use a wavelet transform of L\'evy stable noise
and then reconstruct a fractional L\'evy noise as in the
generation of fBm with wavelet transform \cite{Fland1992,
Stoks1994, Heneg1996, HuZhu2000}. We use the Daubechies wavelet in
the simulations. Thus, fractional L\'evy noise is the series of
increments of the fLm. There are other generating methods, for
instance using Fourier transforms, but the errors are more
difficult to control.

\subsubsection{Numerical simulations}

We simulated fractional L\'{e}vy noises for different Hurst
exponents $H$ and heavy-tailness $\alpha$. For each simulation, we
fixed the pair $(H,\alpha)$ among $19 \times 40$ possible values:
$H$ ranged from $0.05$ to $0.95$ with spacing $0.05$, while
$\alpha$ varied from $0.05$ to $2$ with step $0.05$. Four typical
noises are illustrated in Fig. \ref{Fig:NoisefLn}: (a) $H=0.2$ and
$\alpha=0.4$, (b) $H=0.4$ and $\alpha=0.8$, (c) $H=0.6$ and
$\alpha=1.2$, (d) $H=0.8$ and $\alpha=1.6$.

To qualify the effect of the interplay between correlations and
heavy tails in fractional L\'{e}vy noises, we introduce a
characteristic quantity $h_p$ defined as the Lomb peak height
corresponding to a certain false-alarm probability $p$. We checked
that different choice of $p$ resulted in the same properties of
$h_p$ reported below, as long as $p$ is not too small. We
generated 1,000 time series each with 100 data points for each
couple $(H,\alpha)$ of the $19 \times 40$ possible couples.

Fig. \ref{Fig:hp0.01} shows the dependence of $h_{0.01}$ as a
function of $H$ and $\alpha$ in the fractional L\'evy noise.
$h_{0.01}$ increases both with $H$ and $\alpha$. This is
consistent with previous results obtained for each of the
ingredients: a large correlation (larger $H$) increases the Lomb
peaks (and thus false-alarm probability); a large power law
exponent was also found to increase the false-alarm probability
since this corresponds to weaker fluctuations and thus allows for
the random appearance of coherent periodic structures.

\section{Joint-distribution of the frequency and height of the
highest peak of the Lomb power spectrum} \label{sec:joint}

Up to now, we have reported the false-alarm probability as a
function of the highest Lomb peak, {\it independently} of its
frequency. In practice, the value of the false-alarm probability
as a function of frequency is also an important determinant of the
statistical significance of a supposed periodic signal. The
confidence level of a Lomb peak in a given time series may be
increased if one can distinguish the frequency $f^r$ corresponding
to the highest Lomb peak $h^r$ of the real signal from the most
probable frequency $f^{mp}$ of a noise determined from synthetic
time series. This approach has been used for specific cases in
\cite{Huang2000a,Huang2000b}.

To address this question, we now investigate the impact of the
heavy-tailness of noise distributions and noise correlation on the
two-point statistics of $(f, h)$. We study L\'evy stable noise and
fractional Gaussian noise separately, as this is enough to show
the main facts and the numerical calculations become prohibitive
for the combined fractional L\'evy noises. For each type of noise,
we synthesize 50,000 series each with 100 data points. The same
algorithms as in Sec.~\ref{subsec:LSN} are used for the L\'evy
stable noise and as in Sec.~\ref{subsec:fGn} for the fractional
Gaussian noise.

Fig.~\ref{Fig:PfhStab} shows the probability density distribution
$p(f,h)$ of the highest Lomb peak height $h$ and its associated
frequency $f$ for the L\'evy stable noises, for $\alpha = 0.1,~ 1$
and $1.5$. In this case of uncorrelated L\'evy noise, all
frequencies are equiprobable: this is clearly retrieved in the
distribution for $\alpha = 1.5$. Large fluctuations for the
smaller $\alpha$'s show that the statistical ensemble of these
simulations is not sufficiently large to reach the asymptotic
independence over the frequencies, as large bursts of probability
$p(f,h)$ occur at certain frequencies. The dependence of $p(f,h)$
as a function of $h$ recovers our previous results shown in
Fig.~\ref{LevySigLevel}.

\begin{figure}
\begin{center}
\epsfig{file=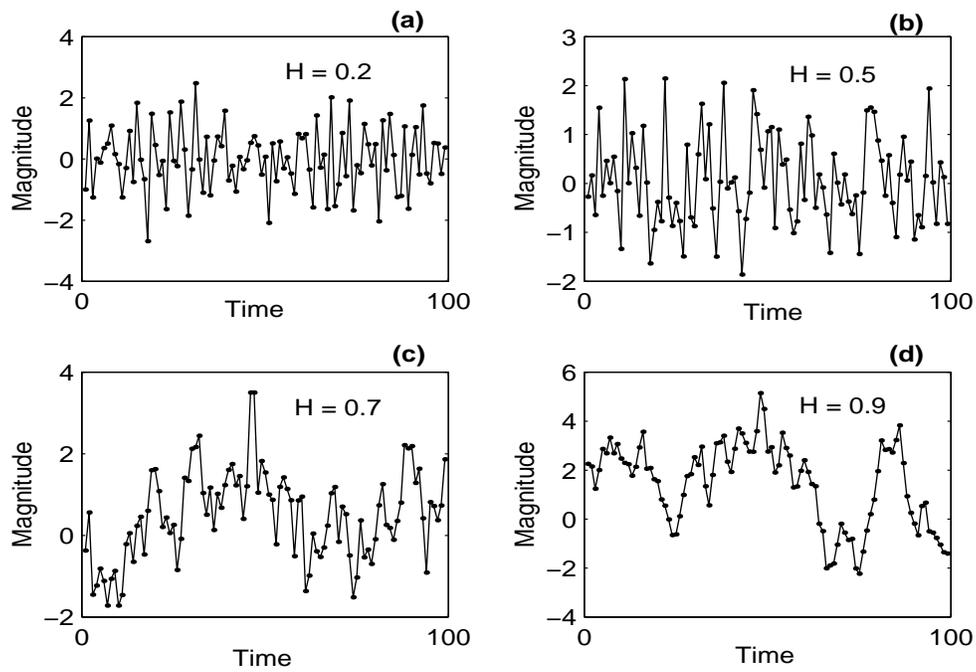,width=13cm, height=9cm}
\end{center}
\caption{\small{Surrogate fractional Gaussian noises using the
Durbin-Levinson algorithm. Each noise has 100 data points whose
Hurst indexes are: (a) $H = 0.2$, (b) $H = 0.5$, (c) $H = 0.7$,
and (d) $H = 0.9$. With the increase of $H$, the fGn exhibits
stronger regularity.}} \label{Fig:NoisefGn}
\end{figure}

\begin{figure}
\begin{center}
\epsfig{file=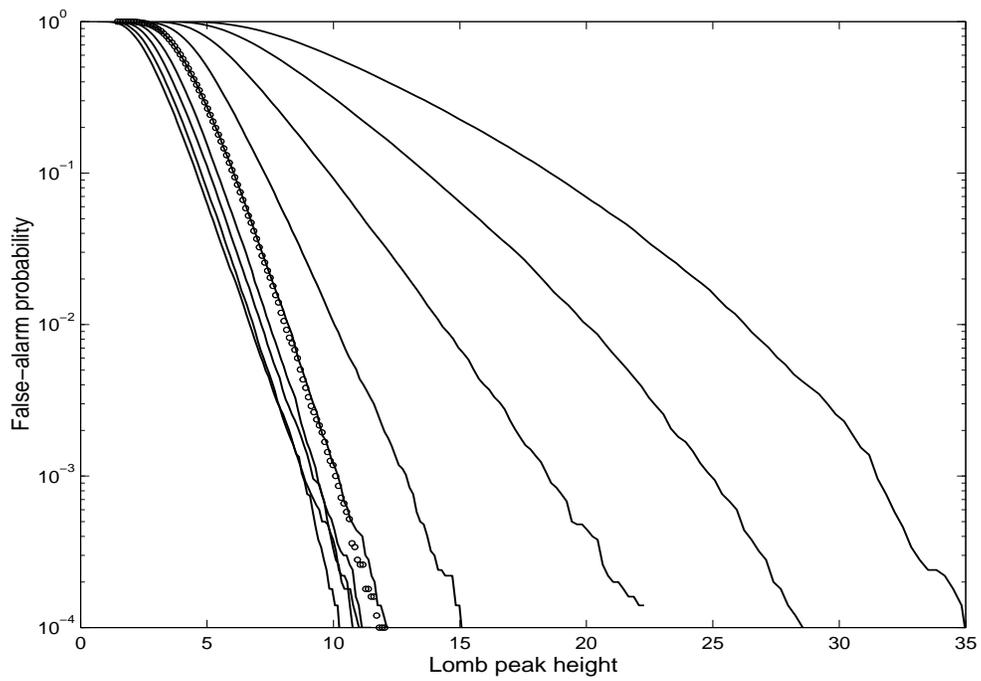,width=13cm, height=9cm}
\end{center}
\caption{\small{The relationship between the Lomb peak height and
the significant level of 50,000 synthetic noises with 100 data points. The
Hurst index increases from 0.1 to 0.9 with spacing 0.1 from left
to right.}} \label{FigLombfGn}
\end{figure}

\begin{figure}
\begin{center}
\epsfig{file=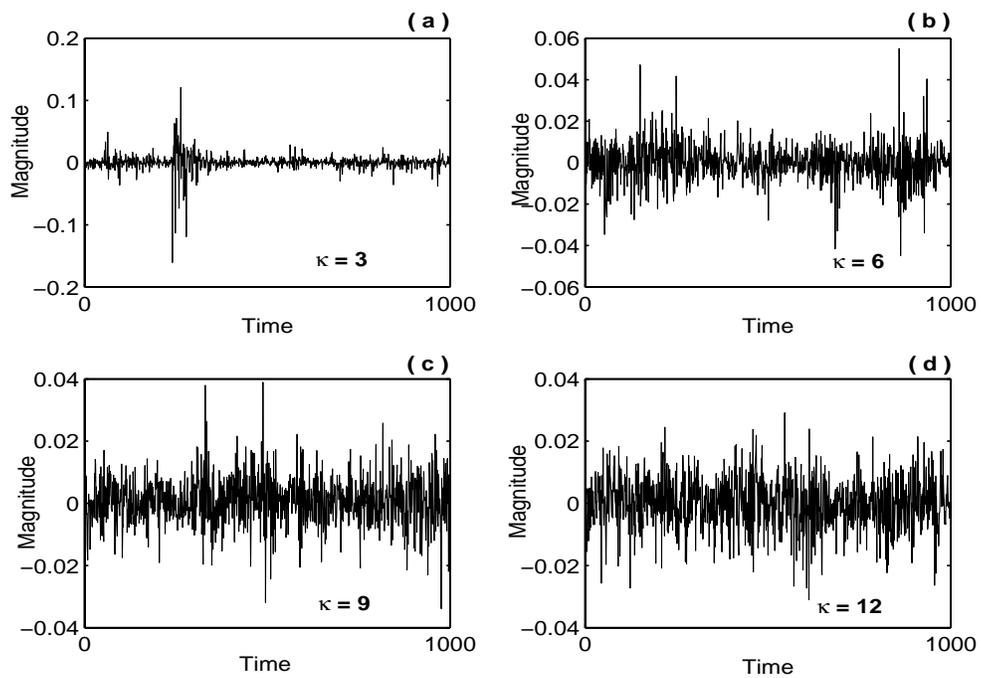,width=13cm, height=9cm}
\end{center}
\caption{Surrogate GARCH(1,1) noises using Eqs. (\ref{Eqe}) with:
(a) $\kappa = 3$, (b) $\kappa = 6$, (c) $\kappa = 9$, and (d)
$\kappa = 12$.} \label{Fig:NoiseGarch}
\end{figure}

\begin{figure}
\begin{center}
\epsfig{file=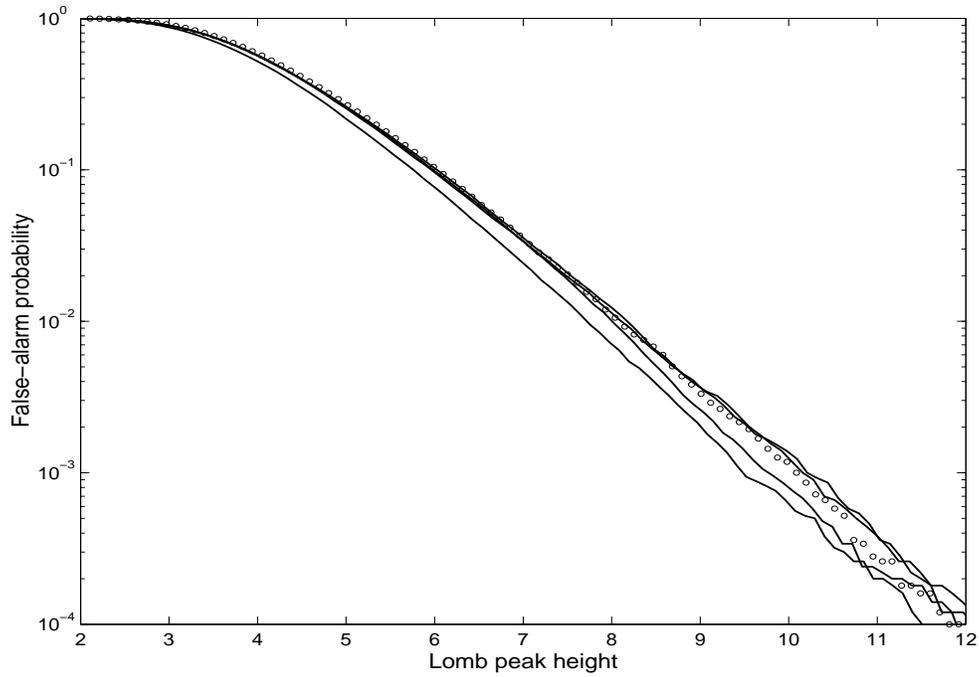,width=13cm, height=9cm}
\end{center}
\caption{\small{Significance levels of the surrogate GARCH(1,1)
data for $\kappa =3, 6, 9, 12$ and comparison with the independent
Gaussian noise.}} \label{FigLombGarch}
\end{figure}

\begin{figure}
\begin{center}
\epsfig{file=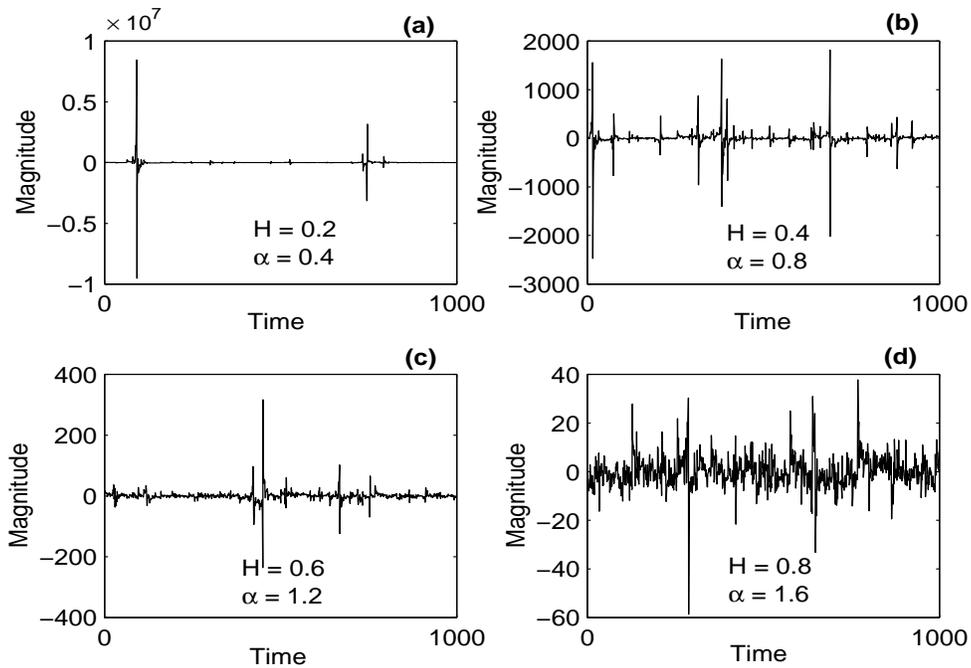,width=13cm, height=9cm}
\end{center}
\caption{\small{Fractional L\'evy noises generated with the
wavelet method using Daubechies wavelets.}} \label{Fig:NoisefLn}
\end{figure}

\begin{figure}
\begin{center}
\epsfig{file=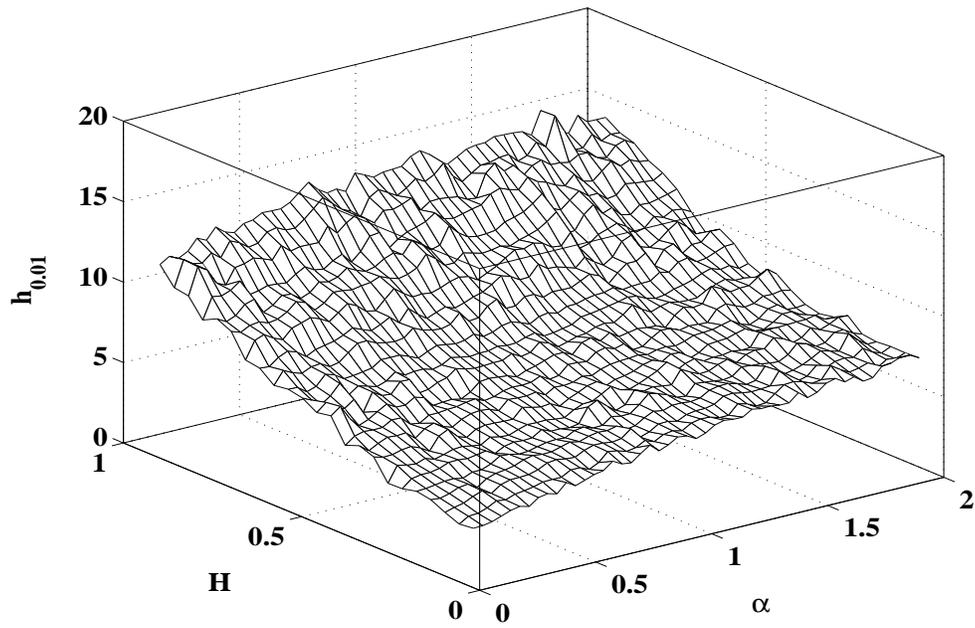,width=13cm, height=9cm}
\end{center}
\caption{\small{Dependence of $h_{0.01}$ on $H$ and $\alpha$ in
fractional l\'evy noise. $h_{0.01}$ increases with $H$
(respectively $\alpha$) for fixed $\alpha$ (respectively $H$).
Two-dimensional interpolations were carried out to smoothen the
surface. The residual oscillations stem from statistical
fluctuations.}} \label{Fig:hp0.01}
\end{figure}

\begin{figure}
\begin{center}
\epsfig{file=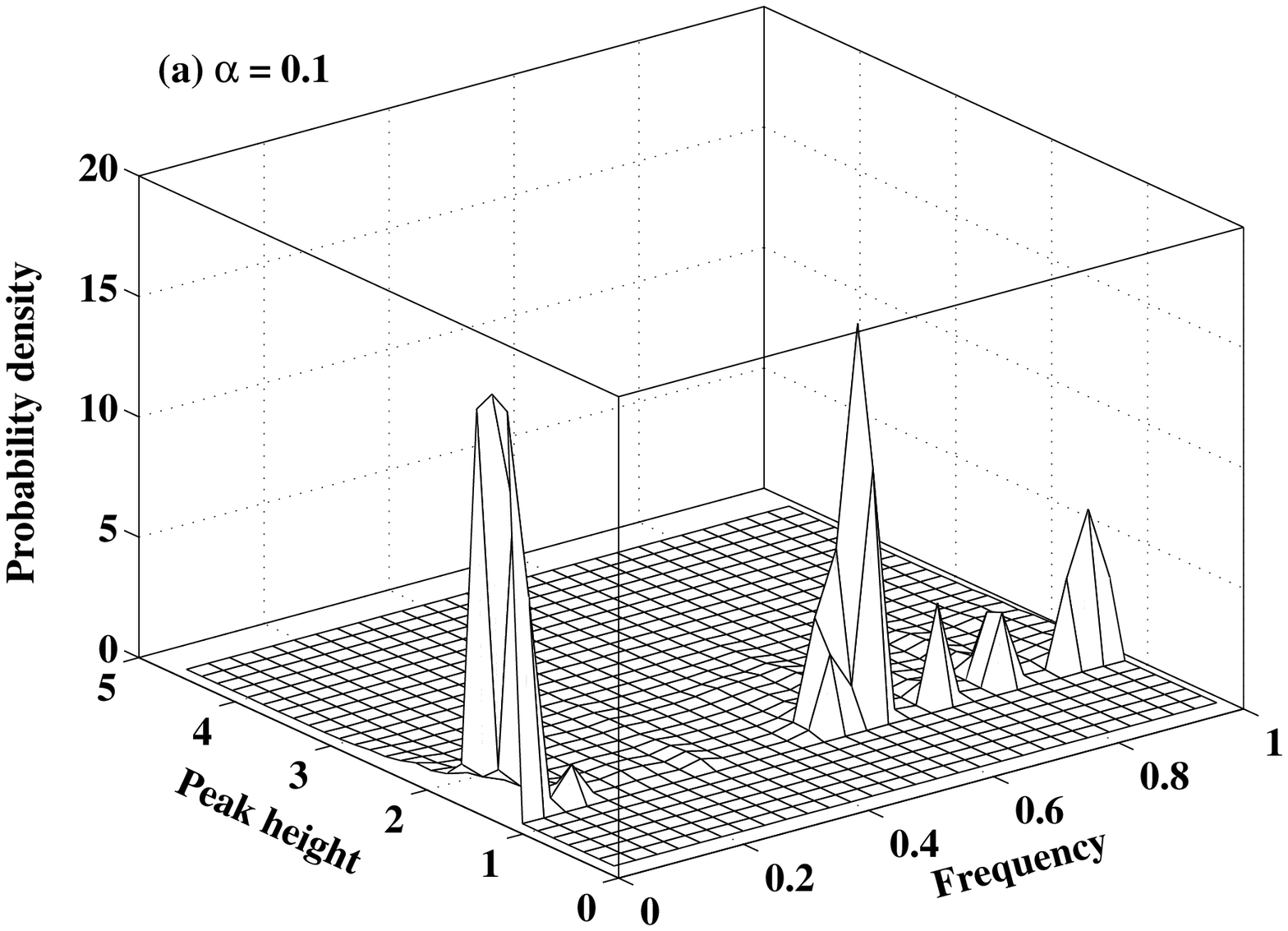,width=9cm, height=6cm}
\epsfig{file=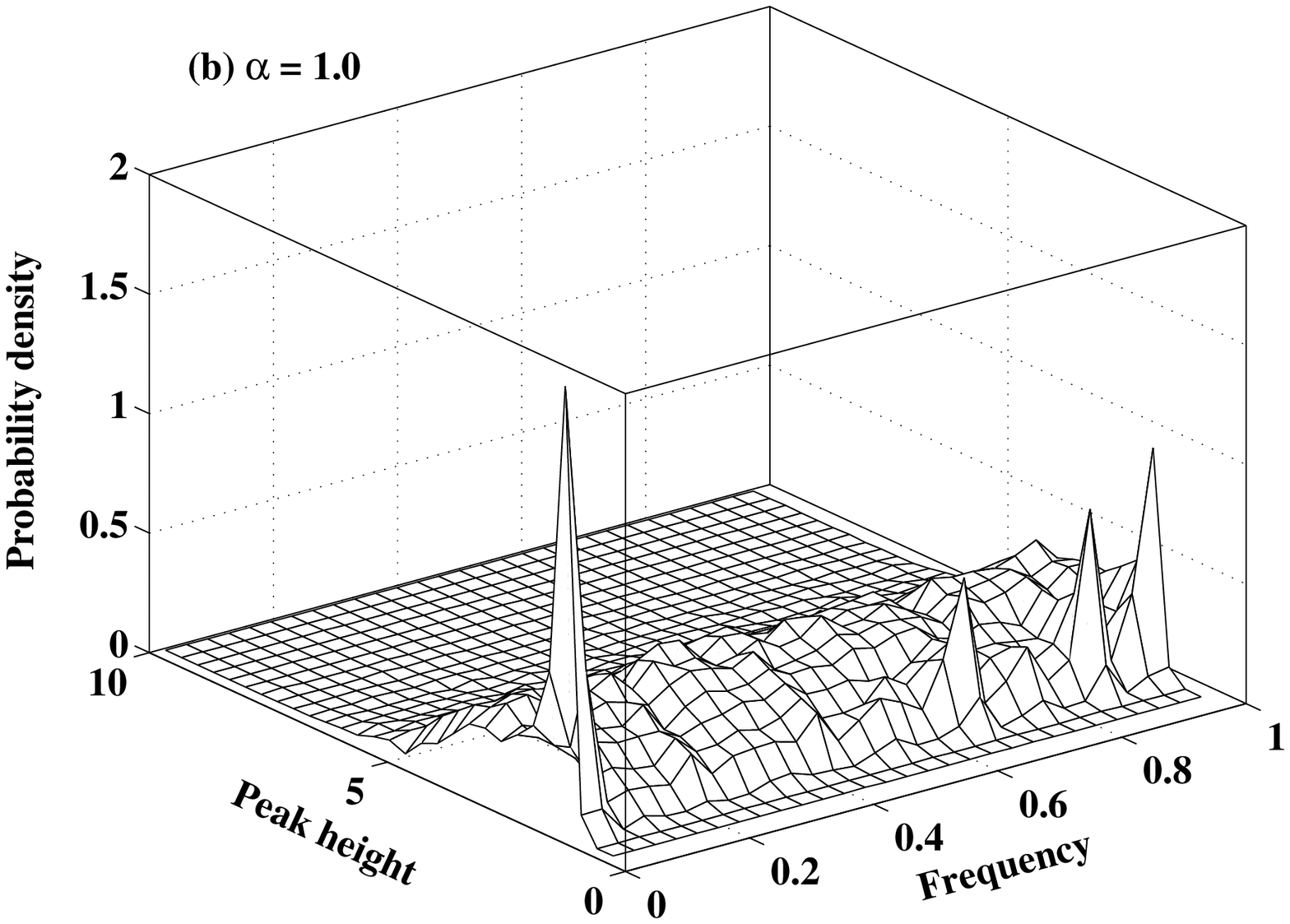,width=9cm, height=6cm}
\epsfig{file=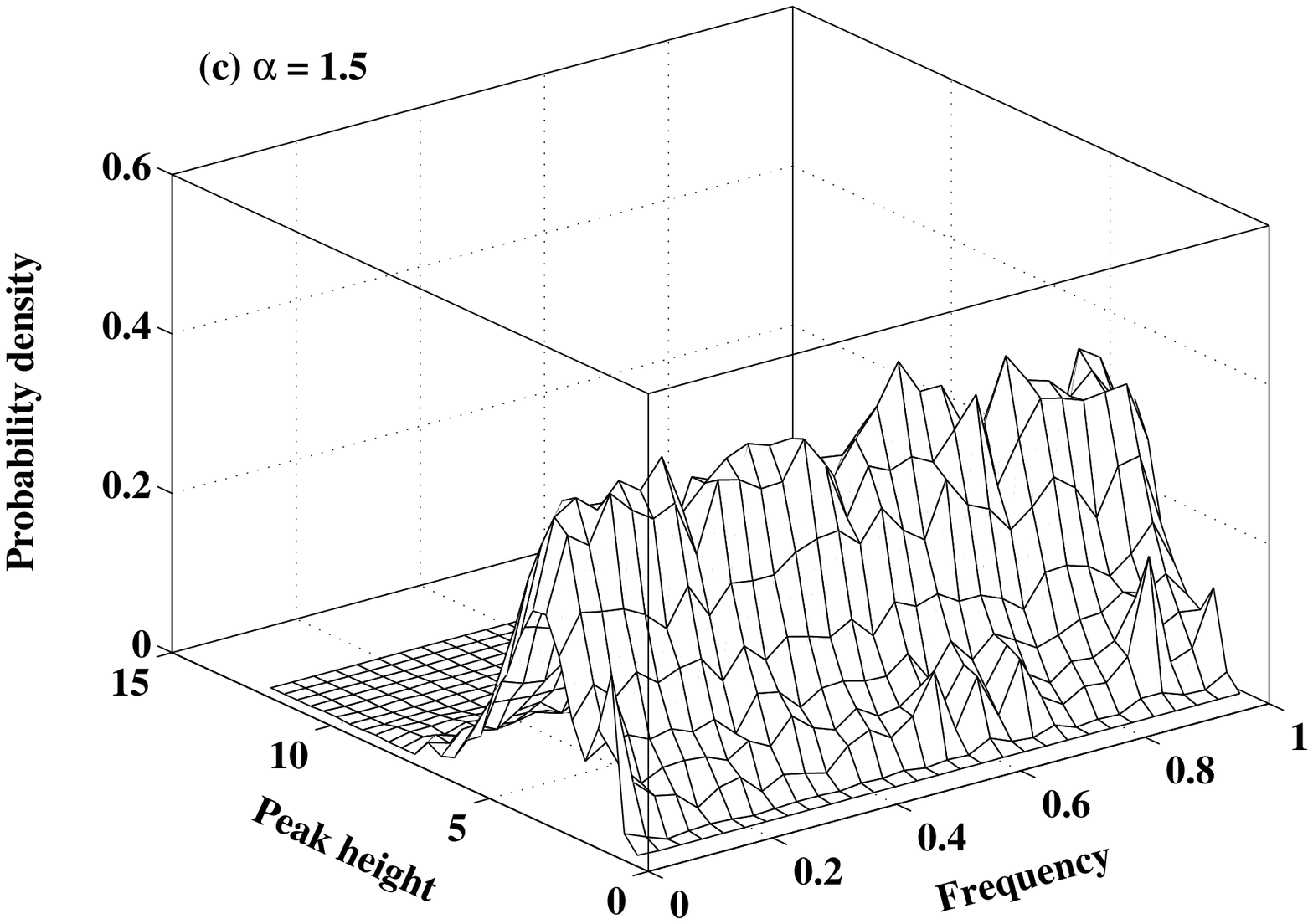,width=9cm, height=6cm}
\end{center}
\caption{\small{Probability density $p(f,h)$ of the highest Lomb
peak height $h$ and its associated frequency $f$ for L\'evy stable
noises: (a) $\alpha = 0.1$, (b) $\alpha = 1$, and (c) $\alpha =
1.5$. }} \label{Fig:PfhStab}
\end{figure}

\begin{figure}
\begin{center}
\epsfig{file=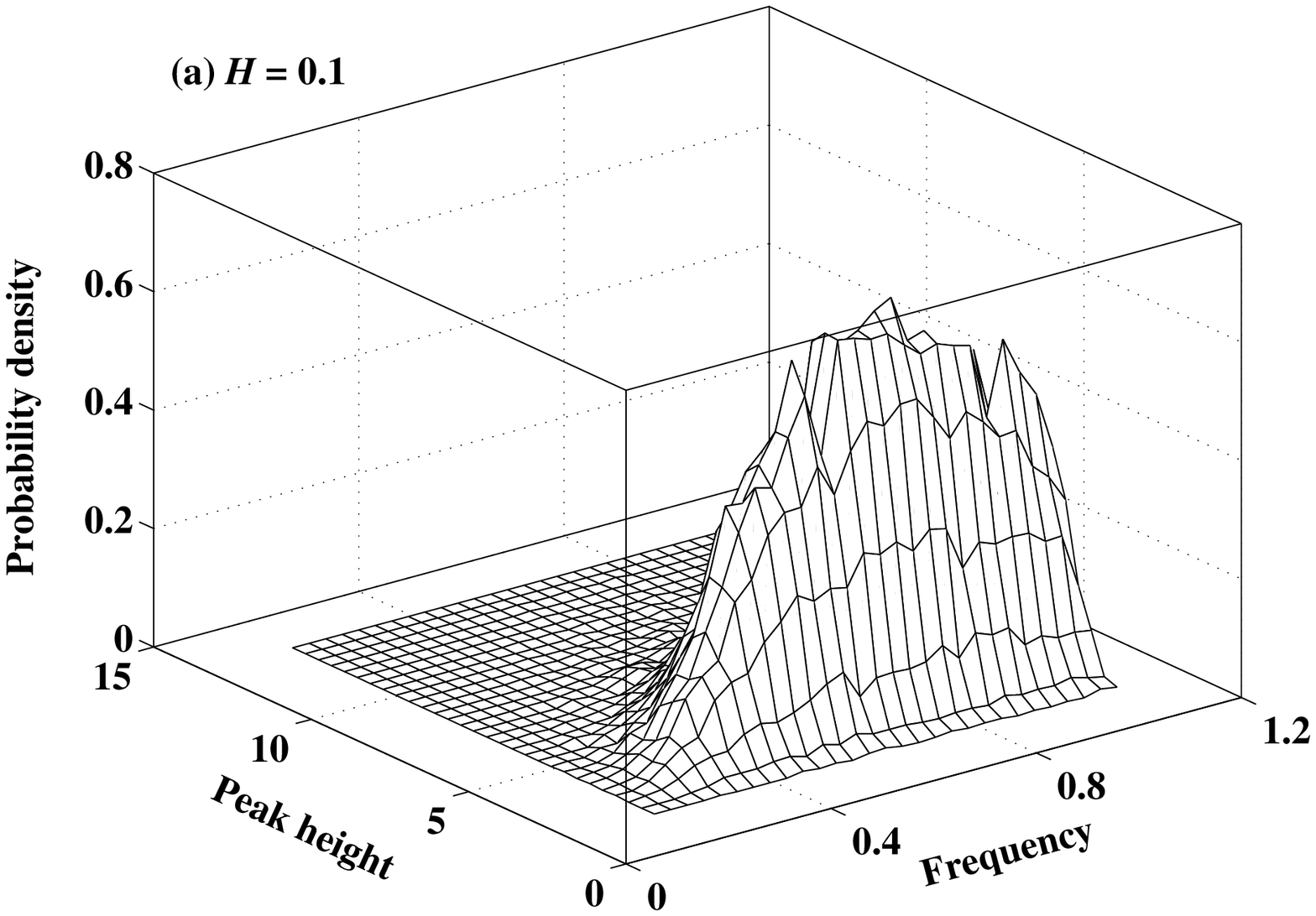,width=9cm, height=6cm}
\epsfig{file=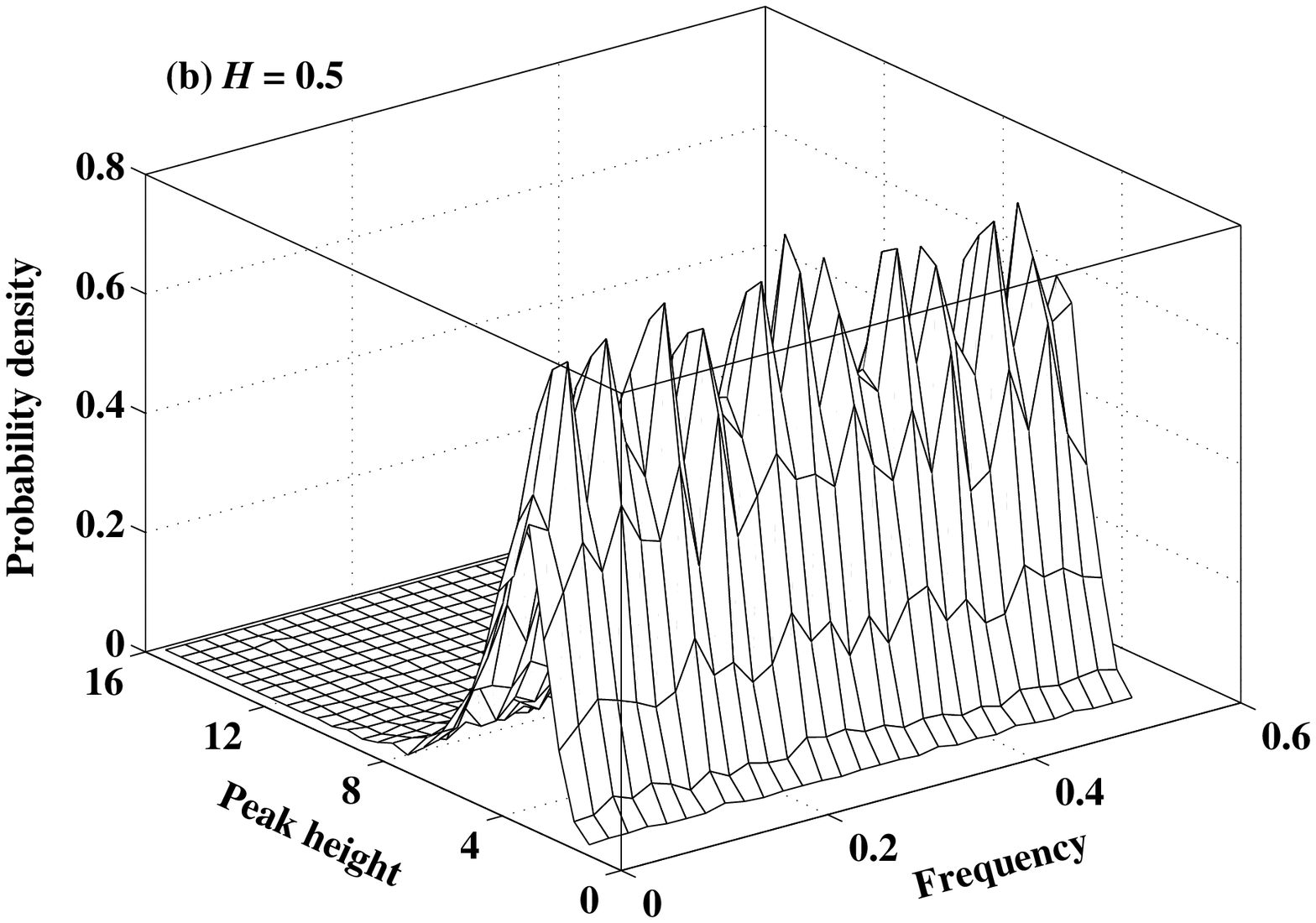,width=9cm, height=6cm}
\epsfig{file=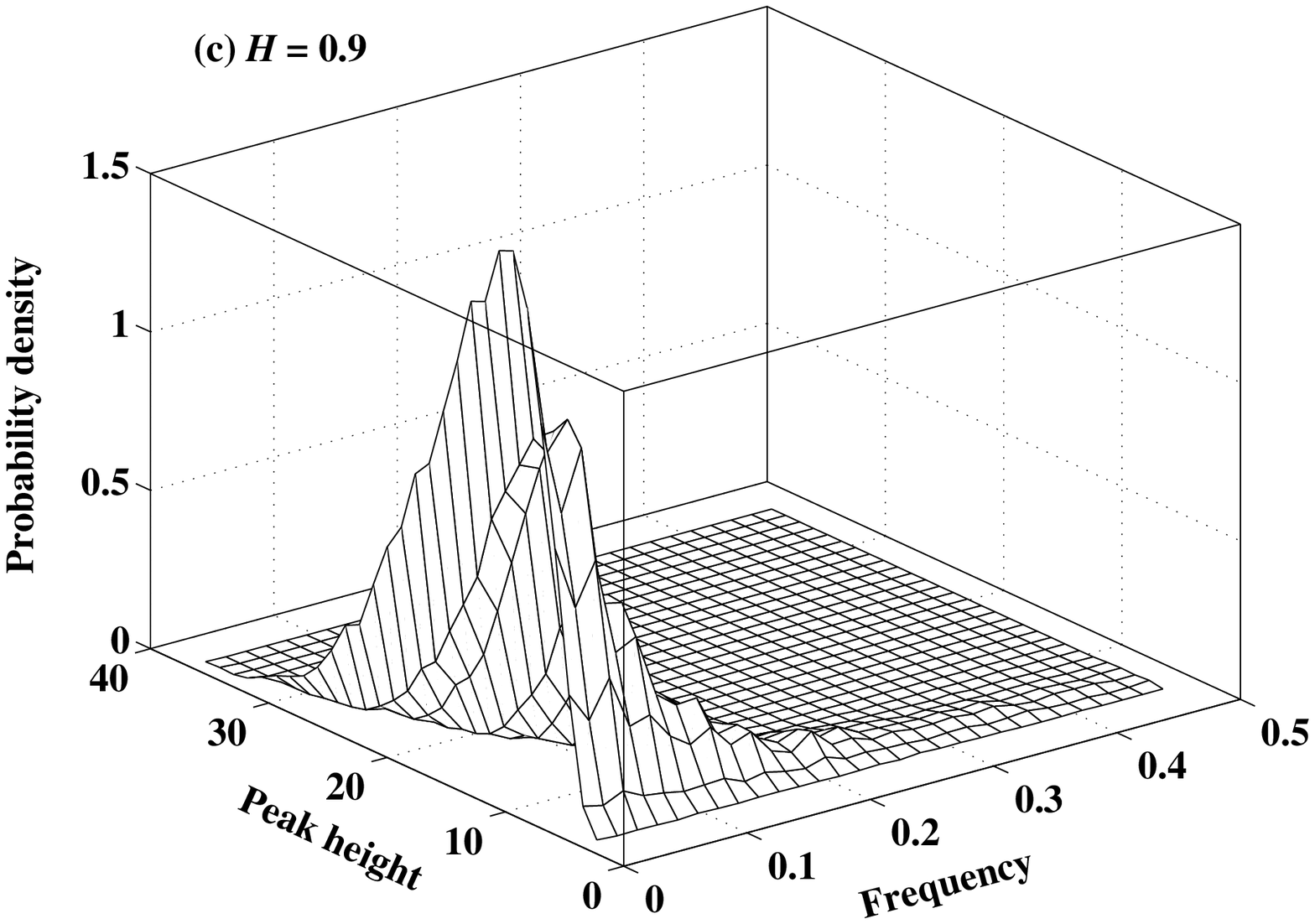,width=9cm, height=6cm}
\end{center}
\caption{\small{Three types of probability density $p(f,h)$ of the
highest peak for fractional Gaussian noise: (a) $H = 0.1$, (b) $H
= 0.5$, and (c) $H = 0.9$.}} \label{Fig:PfhfGn}
\end{figure}

The probability densities $p(f,h)$ of the highest Lomb
peak height $h$ and its associated frequency $f$ for fractional
Gaussian noises are shown in Fig.~\ref{Fig:PfhfGn}, for $H=0.1,
~0.5$ and $0.9$.
\begin{itemize}
\item  For $H=0.1$ and more generally for $0 < H < 0.5$
(Fig.~\ref{Fig:PfhfGn}a), the fGn is anti-persistent, i.e., the
noise tends to reverse its sign, and thus oscillate fast. We
should thus expect and do observe periodic components with high
frequencies. For instance, in Fig.~\ref{Fig:NoisefGn}a generated
for $H = 0.2$, one can observe about $30$ ``cycles'' in the noisy
time series, resulting in $f^{mp} \simeq 0.6$.

\item For $H = 0.5$ (Fig.~\ref{Fig:PfhfGn}b), the fGn is
white-noise and all frequencies should be equivalent, as observed.

\item For $H=0.9$ and more generally for $0.5 < H < 1$
(Fig.~\ref{Fig:PfhfGn}c), the fGn is persistent, i.e., the noise
tends to continue along a trend. We should thus expect and do
observe periodic components with very low frequencies. For
instance, in Fig.~\ref{Fig:NoisefGn}d generated for $H = 0.9$, one
can observe about $2$ ``cycles'' in the noisy time series,
resulting in $f^{mp} \simeq 0.02$.
\end{itemize}
Generally, the number of ``cycles'' decreases as $H$ increases.
Thus, the most probable frequency $f^{mp}$ decreases with
increasing $H$, ranging from $\frac{1}{\triangle t}$ to
$\frac{N}{3\triangle t}$.

We have verified in our simulations that the distribution $p(f,h)$
for the $\alpha = 2$ L\'evy stable noise is identical to the
distribution $p(f,h)$ obtained for the fractional Gaussian noise
with $H = 0.5$, as it should. This provides a confirmation of the
validity of the applied algorithms.

\section{Relation with previous works on the detection of
log-periodicity} \label{sec:appl}

We now briefly discuss how the simulations presented here shed
light on previous announcements of the detection of
log-periodicity. To apply the numerical results in this work,
there are several issues that should be addressed in advance.

\subsection{Impact of amplitudes of harmonics}
\label{subsec:ampl}

It is natural that the underlying log-periodic function in the
noisy signal is not the pure cosine shown in Eq.~(\ref{Eq:yt}).
Then we will see harmonics in the Lomb periodogram. There are
several different cases.

If the amplitude of the fundamental periodic component dominates,
i.e. $y_1 \gg y_m$ for all $m>1$, $P_N(\omega_1) \gg
P_N(\omega_m)$. Note that $\omega_m/\omega_1$ are positive
integers. In this situation, we can apply our numerical results
directly to determine the significance level of the Lomb peak.

In contrast, if the amplitude(s) of the harmonics is comparable
with that of the fundamental periodic component, i.e. $y_1 \simeq
y_m$ for some $m>1$, no less than two comparable peaks coexist in
the periodogram. In this case, it is more difficult to determining
the significance of the peaks, as is implied by
Eq.~(\ref{Eq:PNapprox}).

It is also possible that for some systems there exists at least
one harmonic component $m$ such that $y_1 \ll y_m$ leading to
$P_N(\omega_1) \ll P_N(\omega_m)$. Examples, which we have studied
recently will be reported elsewhere. They include
(1) the log-periodic residues of moments of some self-similar
multinomial measures \cite{Zhou2001a}; (2) $q$-derivative
\cite{Erzan1997} of some ``Weierstrass-type'' functions
\cite{Simon2001}; and (3) canonically averaged local
log-derivative of the moments of energy dissipation rate in
three-dimensional fully developed turbulence \cite{Zhou2001b}.

There are other possible origins which might cause several high
peaks in one Lomb periodogram, with low signal-to-noise ratio. If
these peaks are disordered, it is hard to conclude that periodic
component(s) exist. Meanwhile, high noise will also suppress the
fundamental peak and increase harmonic peaks due to the
interaction between signal and noise.

For the situations that $y_1 \gg y_m$ fails, all the
peaks are not significant if we apply the numerical results
directly. However, it is well known that evenly spaced high
peaks in the power periodogram is a strong signature of the
existence of a fundamental frequency. We can use a least-square
fitting to extract the fundamental frequency as proposed in
Ref.~\cite{Zhou2001a, Zhou2001b}. This is however beyond the theme
of the present paper.

\subsection{Statistics of highest peaks ratio}
\label{subsec:peakratio}

\begin{figure}
\begin{center}
\epsfig{file=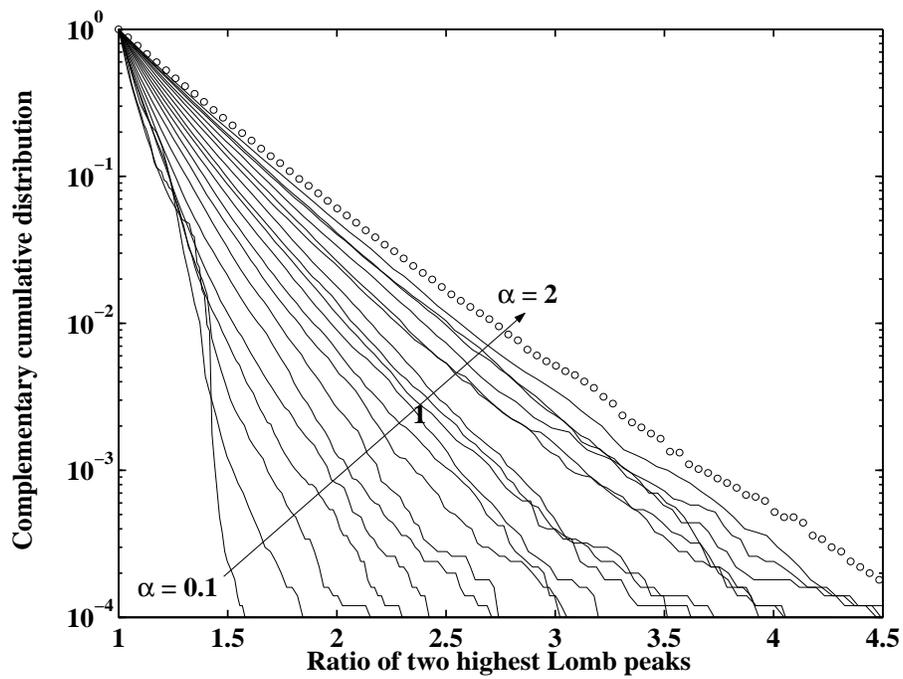,width=12cm, height=9cm}
\end{center}
\caption{\small{Complementary cumulative distribution of the ratio
of the highest Lomb peaks for L\'evy stable noise with $\alpha$
varying from 0.1 to 2 as indicated by the arrow. The open circles
corresponds to $\alpha = 2$.}} \label{Fig:iLnPeakRatio}
\end{figure}

In practice, it is always difficult to judge if a given Lomb peak
is significant or not, especially when the nature of the noise is
unknown. Then, the absolute amplitude of the highest Lomb peak is
difficult to interpret and translate into a false-alarm
probability and into a confidence level. It is then natural to
introduce a measure of relative significance, for instance by
using the ratio of the first highest Lomb peak to the second
highest peak. Fig.~\ref{Fig:iLnPeakRatio} and
Fig.~\ref{Fig:fGnPeakRatio} show the complementary cumulative
distributions of the ratio of the two highest Lomb peaks for
L\'evy stable noise and fractional Gaussian noise. Again, the
number of samples is 50,000 each with 100 data points. The open
circles in the two figures correspond to Gaussian noise and the
slight discrepancy is caused by the application of different
algorithms. The complementary cumulative (false-alarm) probability
of a given peak increases with $\alpha$ for L\'evy stable noise,
which corroborates our previous statistics on the absolute height
of Lomb peaks: for small exponent $\alpha$, it is very improbable
to observe a large ratio of the two highest Lomb peaks. Thus,
a signal showing a large ratio would qualify as highly
significant.

The behavior of fGn is non-monotonous. The false-alarm
(complementary cumulative) probability decreases with $H$ in the
``anti-persistence'' regime $0 < H \leq 0.5$ and increases with
$H$ in the persistence regime $0.5 \leq H <1$. Thus, both
mechanisms, which leads to deviations from uncorrelated
randomness, lead to an increase the false-alarm probability.

\subsection{Application to real systems}
\label{subsec:appl}

We now review how the present analysis
shed light on our previous works that have used the Lomb analysis
to attempt qualifying the presence of log-periodicity in different
complex systems.

Ref.~\cite{Huangpreneedle} presents Lomb periodograms of the
residual of the fit of the distribution of crack lengths by a pure
power law. Since the fit is performed with a power law, the
natural variable is the logarithm $\ln \ell$ of the crack lengths:
a detection of a periodic component in the residual of the
distribution of the crack lengths as a function of $\ln \ell$
would qualify log-periodicity. The Lomb periodogram shown in Fig.~14
of \cite{Huangpreneedle}
for a model of crack growth has a very
high peak of $90$ with very narrow width and is highly
significant, whatever the nature of the noise, Gaussian, L\'evy,
power law or even with long-range correlations.
The Lomb periodogram shown in Fig.~16 of \cite{Huangpreneedle}
for geological data on the distribution of joint length has two peaks
at a level between $45$ and $50$ which are so close as being barely
distinguishable. All the other peaks are much smaller not above 13. 
These  two peaks
are very significant, even in the presence of
the most unfavorable case of highly persistent noise.

Ref.~\cite{Johsorhanturb} analyzed experimental data on the time
evolution of the number, size and separation of vortices in freely
decaying 2-d turbulence to investigate the existence of a {\it
discrete} time scale invariance, which could reflect that the
time-evolution of the merging of vortices is not smooth but
punctuated, leading to a preferred scale factor and as a
consequence to log-periodic oscillations. Three Lomb periodograms
were reported which were averaged over 7-10 realizations of an
experiment on freely decaying turbulence: Figs.~5 and 6 of
Ref.~\cite{Johsorhanturb} show the Lomb periodogram of the
logarithmic derivative of the number of vortices with respect to
time; Figs.~9 and 10 show the Lomb periodogram of the logarithmic
derivative of the separation between vortices with respect to
time; Figs.~13 and 14 show the Lomb periodogram of the logarithmic
derivative of the mean radius of vortices with respect to time. In
all cases, the maximum height is no more than about $3.5$. The
present study suggests that only for heavy-tailed noise with small
exponent $\alpha$ or $\kappa$ will each Lomb periodogram achieve a
reasonable statistical significance.

In \cite{Johsorhanturb}, it was argued
that the case for log-periodicity was not based on the evidence
obtained for each isolated data set but from the collective evidence that
the maximum peaks were found for each data set at about the same log-frequency
around $4-5$ corresponding to a preferred scaling ratio $1.3$. This
log-frequency of $4-5$
is much higher than the frequency one would expect from the size
of the interval (see Fig.~4 of \cite{Johsorhanturb} and the value of the
most probable frequency due to noise equal to $1.5/1.8=0.8$)
Thus, statistical significance is argued to be
based on the coincidence of three highest peaks of the three Lomb
periodograms which occur at frequencies far from the most probable
frequencies found in our studies. Taking a number of points around
$35$ in these data, the signal-to-noise ratio is around
$y_m^2/2\sigma^2_m = 0.2$ according to Eq.~(\ref{Eq:PNapprox}).
This might be the reason for the low Lomb peaks in the periodograms. On the
other hand, we cannot find support from the statistics proposed in
Sec.~\ref{subsec:peakratio}. A further note of caution is
necessary: the curves in figures 4, 8, and 12 of
Ref.~\cite{Johsorhanturb} are quite similar to fractional Gaussian
noise with  $H \approx 0.3$. According to this alternative
interpretation, long-range correlations could characterize the
this turbulence data, rather than log-periodicity. This example shows the
difficulty and complication
of determination of the significance level of a given peak in
practice.

Ref.~\cite{kobenew} reports in its Fig.~1 five Lomb periodograms
of five time-dependent ion concentrations in water close to the
epicenter of the Kobe earthquake of Jan. 17, 1995 as a function of
log-frequency. The two largest peaks are of height $13$ and $11$.
According to the uncorrelated Gaussian benchmark, these two peaks
give a false-alarm probability of about or less than $10^{-4}$,
suggesting a very strong significance for a genuine
log-periodicity. The possible existence of tail-tailed noise
reinforce even further this case. However, one cannot exclude the
possible existence of some long-range correlation in the ion
concentrations, for instance associated with the time-dependent
evolution of water permeability. Our present study shows that
going from a white-noise spectrum to a $1/f$ spectrum transform
the false-alarm probability from $10^{-4}$ to about $10-20\%$.

\begin{figure}
\begin{center}
\epsfig{file=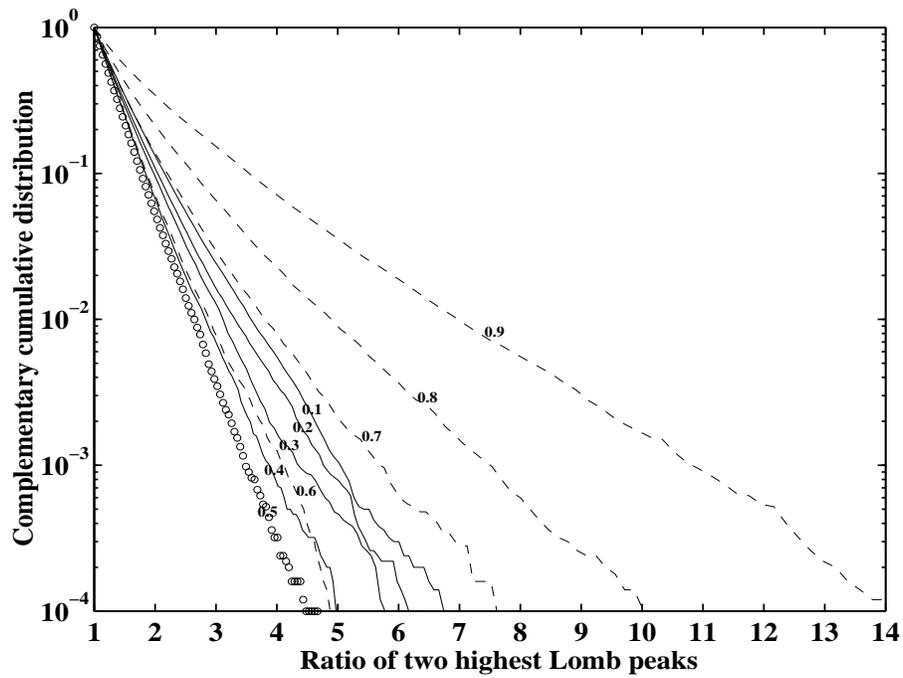,width=12cm, height=9cm}
\end{center}
\caption{\small{Complementary cumulative (false-alarm) distribution
of the ratio
of the highest Lomb peaks for   fractional Gaussian noise with $H$
varying from 0.1 to 0.9. The open circles, solid lines and dashed
lines correspond respectively to $H = 0.5$, $0 < H < 0.5$ and $0.5
< H <1$. }} \label{Fig:fGnPeakRatio}
\end{figure}

References \cite{johsor2, JSantibubble, johsor3, emergentmark,
signifsorjan} report several Lomb periodograms performed to
qualify the existence of log-periodicity in the price dynamics of
speculative bubbles preceding large financial crashes. The Lomb
power spectra are normalized such that the highest peak has a
height equal to $1$. Therefore, we cannot establish the
corresponding false-alarm probability, using the different models
of noises investigated here. It was argued that the correct
method, in absence of information on the underlying noise, is the
establish the statistical significance on the basis of a large
difference between the highest peak and the background. Figure
\ref{Fig:fGnPeakRatio} shows that a white-noise null hypothesis
requires a ratio larger than about $3$ to qualify a genuine
log-periodic component at the $99\%$ confidence level. In the
worst and unrealistic scenario where the noise exhibits long-range
correlations with persistence, the ratio must be larger. For
instance, figure \ref{Fig:fGnPeakRatio} indicates that the ratio
must reach $5$ if the Hurst exponent is 0.8 or smaller. However,
financial time series are known to have very short-ranged and weak
correlations in their return and it is thus hard to believe that
strong persistence with large Hurst exponents can characterize the
statistics of the returns. We stress here that we are not
referring to the persistence property of the absolute value of the
returns (or volatility) but to that of the returns themselves
which is an entirely different problem. Many of the ratios of the
larger peak to the second largest one reported in the Lomb
periodograms of \cite{johsor2, JSantibubble, johsor3,
emergentmark, signifsorjan} are of the order or larger than 3,
some reach 5 or more. This signals a high statistical
significance. In some instance, the ratio is not large, but it
turns out that the second highest peak is associated with a
log-frequency harmonic to the log-frequency of the first peak. In
this case, the statistics of the ratio does not apply as the
existence of harmonics supports the conclusion they are not
generated from noise. Hence, the statistical analysis of the ratio
of the first highest peak to the second one confirms the
confidence of the existence of log-periodicity in
finance crashes.

\section{Conclusions}
\label{sec:conclusion}

We have presented statistical tests on the false-alarm probability
that a spectral peak in the Lomb periodogram analysis of noisy
periodic signal may result from noise. In order to mimic the large
variety of noises that may be present in natural and social data,
we have investigated several types of noises, including noises
with power law distributions and with short and long-range
correlations. The false-alarm probability of a periodic component
is found to be strongly dependent on the nature of the noise. This
underlines the difficulty in concluding unambiguously on the
existence of a genuine log-periodicity in noisy signals when the
noise properties are not known a priori and are thus difficult to
distinguish from the signal. In the light of the statistical tests
performed here, we have briefly reviewed the evidences presented
in past works on the existence of log-periodicity in
turbulence, earthquake and financial data.
Our present study weakens the credibility for 2D-freely decaying turbulence and
strengthens it for fracture, for the ion-signature precursors to the 
Kobe earthquake
and for financial markets.

We hope that the present work will help in assessing more reliably
the existence of periodicity in noisy complex time series and will
provide useful guidelines to test new data sets.

\bigskip
{\bf Acknowledgments:} We thank T. Huillet and J.-F. Muzy for
discussions and D. Stauffer for a careful reading of the manuscript.
This work was partially supported by NSF-DMR99-71475
and the James S. Mc Donnell Foundation 21st century scientist
award/studying complex system.

%\pagebreak

\pagebreak
           %%%%%%%%  Levy stable noise  %%%%%%%%%%
%FIGURE 1

%FIGURE 2

%FIGURE 3

           %%%%%%%%  Power-law noise  %%%%%%%%%%
%FIGURE 4

%FIGURE 5

%FIGURE 6

%FIGURE 7

           %%%%%%%%  Fractional Gaussian noise  %%%%%%%%%%
%FIGURE 8

%FIGURE 9

             %%%%%%%%  GARCH(1,1) noises  %%%%%%%%%%%%%%%
%FIGURE 10

%FIGURE 11

           %%%%%%%%  Fractional Levy noise  %%%%%%%%%%
%FIGURE 12

%FIGURE 13

%%%%%%%%%%%%%%%%%%%%%%p(f, h)%%%%%%%%%%%%%%%%%%%%%%%
%FIGURE 14

%FIGURE 15

%%%%%%%%%%%%%%%%%%%%%%Highest peaks ratio%%%%%%%%%%%%%%%%%%%%%%%
%FIGURE 16

%FIGURE 17

\end{document}